\documentclass{aa}  

\usepackage{graphicx}
\usepackage{txfonts}
\usepackage{natbib}
\usepackage{xspace}
\usepackage{color}
\usepackage{hyperref}

\usepackage{threeparttable}
\usepackage{supertabular}
\usepackage{multirow}
\usepackage{makecell}

\usepackage{amsmath}
\usepackage{subcaption} 
\usepackage{upgreek}
\usepackage{float}

\usepackage{tablefootnote}
\usepackage{tikz}

\newcommand{\cref}{\ref}

\usepackage[normalem]{ulem}

\begin{document}


\title{Timing and spectral variability in 2S 1417-624 observed with Insight-HXMT}

\author{Qi Liu\inst{1}
\and Lingda Kong\inst{1}
\and Can Güngör\inst{2}
\and Lorenzo Ducci\inst{1}
\and Long Ji\inst{3}
\and Wei Wang\inst{4}
\and Xiaohang Dai\inst{1}
\and Andrea Santangelo\inst{1}
}
\institute{Institut für Astronomie und Astrophysik, Universität Tübingen, Sand 1, 72076 Tübingen, Germany \\
\email{qi.liu@mnf.uni-tuebingen.de}
\and
İstanbul University, Science Faculty, Department of Astronomy and Space Sciences, Beyaz{\i}t, 34119, İstanbul, Turkey 
\and 
School of Physics and Astronomy, Sun Yat-Sen University, Zhuhai 519082, China
\and
Department of Astronomy, School of Physics and Technology, Wuhan University, Wuhan 430072, China 
}
\date{ }

\abstract
{We present the results of the spectral and timing analyses of the accreting X-ray pulsar, 2S 1417-624, during the 2018 and 2021 outbursts with Insight-HXMT. We find that the pulse profiles in all energy bands exhibit clear double-peaked structures at low flux states. In the 1–10 keV band, the pulse profiles evolve from double to triple peaks at a flux level of $\sim$4.1$\ \times \ 10^{-9}$ erg cm$^{-2}$ s$^{-1}$, and from triple to quadruple peaks at $\sim$6.4$\ \times \ 10^{-9}$ erg cm$^{-2}$ s$^{-1}$. In the 10-30 keV and 30-100 keV bands, the pulse profiles become narrower at the first transition flux level, followed by a stark transition to quadruple-peaked and triple-peaked structures around the second flux level, respectively. The change of the pulse profile {during the second transition} reveals the transition of the emission pattern from the sub-critical (pencil beam) to the supercritical (fan beam) regime. By performing the binary orbital fitting of the observed spin periods, we provide new measurements of the orbital parameters from the 2021 outburst. Applying different accretion torque models and using the critical luminosity inferred from the pulse profile transitions, we derive a self-consistent distance of {2S 1417-624} in the range of approximately 12.0-15.0~kpc, based on the magnetic field strength derived from the cyclotron resonance scattering feature (CRSF). From the estimated distance of 13 kpc and Gaia's distance of 7.4 kpc, we can infer the observed transition luminosity of \((1.0-1.4) \times 10^{38} \, \mathrm{erg \, s^{-1}}\) and \((3.0-5.0) \times 10^{37} \, \mathrm{erg \, s^{-1}}\), respectively, and compare them with theoretical models. The spectral continuum parameters and the hardness ratio also show significant transitions around the second transition, strongly supporting a change in the accretion regime.}
      

\maketitle

\section{Introduction}
\label{sec:introduction}

The transient Be/X-ray binary 2S 1417-624 was discovered by the Third Small Astronomy Satellite (SAS-3) in 1978 \citep{Apparao1980}. Its spin period was measured to be approximately 17.64 seconds \citep{1981ApJ...243..251K}. \citet{Finger1996} reported an average spin-up rate of $\sim$(3--6)$~\times~10^{-11}$ Hz s$^{-1}$ and determined the orbital period and eccentricity to be around 42.12 days and 0.446, respectively. The optical companion has been identified as a B1 Ve star, located at a distance of 1.4--11.1 kpc \citep{1984ApJ...276..621G}, with Gaia estimating a distance of $9.9^{+3.1}_{-2.4}$ kpc \citep{Bailer2018} based on Gaia DR2. However, the distance is now $7.4^{+3.1}_{-1.8}$ kpc in Gaia DR3 \citep{2021yCat.1352....0B}. \cite{2020MNRAS.491.1851J} estimate a distance of approximately 20 kpc by accounting for the effects of accretion torque. To date, six Type II outbursts have been observed and studied in this system. These outbursts, which are independent of orbital phase, typically last for a significant portion of the orbital period or even span several orbits, during which the peak luminosity can reach up to $\sim$$10^{38}$ erg s$^{-1}$ \citep{2011Ap&SS.332....1R}.

After its discovery, 2S 1417-624 entered a prolonged quiescent period until the second outburst was detected in 1994 by the Burst and Transient Source Experiment (BATSE) \citep{Finger1996}, followed by a third outburst in 1999, observed by the \textit{Rossi} X-ray Timing Explorer (RXTE) \citep{Inam2004}. The fourth giant outburst, monitored by \textit{Swift}/BAT in 2009, reached a peak flux of approximately 300 mCrab. \citet{2018MNRAS.479.5612G} report that a double-peaked pulse profile at lower luminosities evolve into a triple-peaked structure as luminosity increase, based on RXTE observations. They also find an anti-correlation between pulse fraction and source flux and suggest that the beam pattern might transition from sub-critical to supercritical accretion regimes.

The fifth and brightest giant outburst occurred in 2018, with a peak intensity of approximately 350 mCrab as recorded by \textit{Swift}/BAT in the 15–50 keV energy band (see Fig. \cref{fig:counts}). During this outburst, the energy-resolved pulse profiles showed a four-peaked structure at lower energies, which gradually evolved into a double-peaked profile at higher energies, as observed by the Nuclear Spectroscopic Telescope Array (NuSTAR) \citep{2019MNRAS.490.2458G}. Furthermore, the Neutron Star Interior Composition Explorer (NICER) observations reveal that the pulse profiles of 0.2 to 10.0 keV transition to a quadruple peak structure with increasing luminosity \citep{2020MNRAS.491.1851J}. \cite{2020MNRAS.491.1851J} also used the Hard X-ray Modulation Telescope (Insight-HXMT) observation, with a focus only on the 25-100 keV energy bands over three different intensity states. \cite{Liu_2024} present the detailed evolution of the pulse profiles in a broader energy band in the 2018 giant outburst with Insight-HXMT. Compared to \cite{Liu_2024}, we utilise the same dataset but provide more accurate flux estimates from each spectral fit to determine the transition flux level in which the pulse profile changes. More recently, an outburst in 2021 was detected by \textit{Fermi}/GBM and \textit{Swift}/BAT, and studied by \citet{2022Ap&SS.367..112M}. Insight-HXMT also performed high-cadence observations during the outburst from January 29, 2021, to March 14, 2021, which cover approximately one binary orbit. This allows us to perform new measurements of orbital elements through the timing analysis.

The X-ray spectra of 2S 1417-624 are typically modelled using a cut-off power law (cutoffpl in XSPEC) \citep{2018MNRAS.479.5612G,2019MNRAS.490.2458G,2020MNRAS.491.1851J}, or a high-energy cut-off power-law (highecut in XSPEC) continuum model \citep{2022MNRAS.510.1438S,2022Ap&SS.367..112M}. Additionally, an iron emission line around 6.4 keV can be detected \citep{2018MNRAS.479.5612G,2019MNRAS.490.2458G,Liu_2024}. Chandra observed 2S 1417-624 in its quiescent state in May 2013 and \citet{2017MNRAS.470..126T} concluded that the source spectrum could be modelled by either a power law with the photon index of $\sim$0.6 or a blackbody function with a significantly high temperature of approximately 1.5 keV. Several measurements of the magnetic field strength from the torque model for 2S 1417-624 have been reported to date \citep{2020MNRAS.491.1851J,2022Ap&SS.367..112M}. No detection of cyclotron resonant scattering feature (CRSF) was made in the 0.9–79 keV range during the 2018 giant outburst \citep{2019MNRAS.490.2458G}. However, \cite{Liu_2024} recently discovered a CRSF feature at around 100 keV with high statistical significance during the 2018 peak outburst. The detection of a CRSF is crucial for determining the magnetic field strength and advancing our understanding of accretion processes near the neutron star surface. In this paper, we study a detailed timing and spectral analysis of the X-ray pulsar 2S 1417-624 from all observations in the 2018 and 2021 outbursts using Insight-HXMT and present results on spectral variability for the first time. In Sect. \cref{sec:Data} we provide a brief description of the Insight-HXMT observations and data reduction process. Section \cref{sec:results} contains the timing and spectral analysis. We discuss our findings in Sect. \cref{sec:discussion} and present our conclusions in Sect. \cref{sec:summary}.

\section{Observations} \label{sec:Data}

\begin{figure}
    \centering
    \includegraphics[width=.5\textwidth]{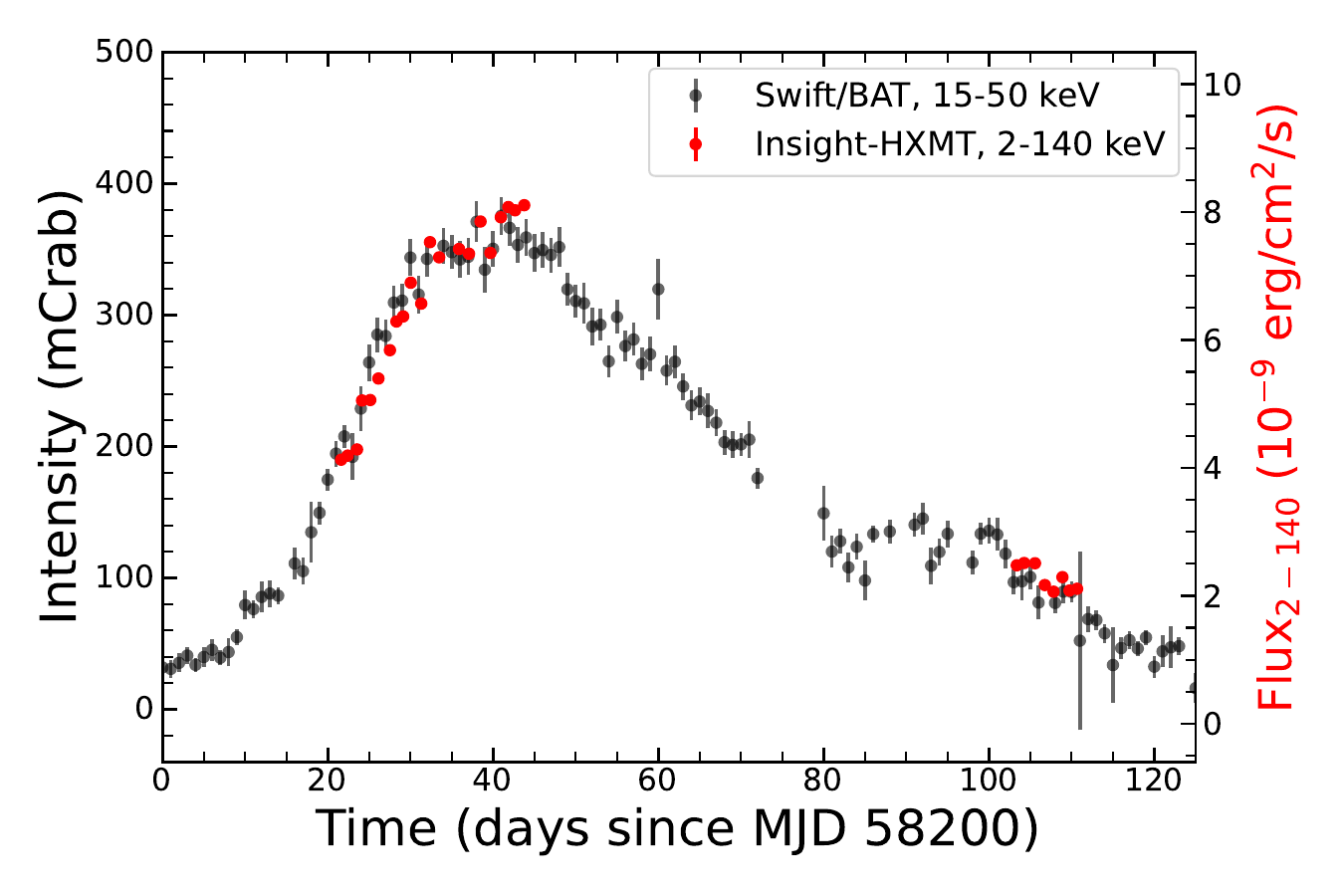}
    \includegraphics[width=.5\textwidth]{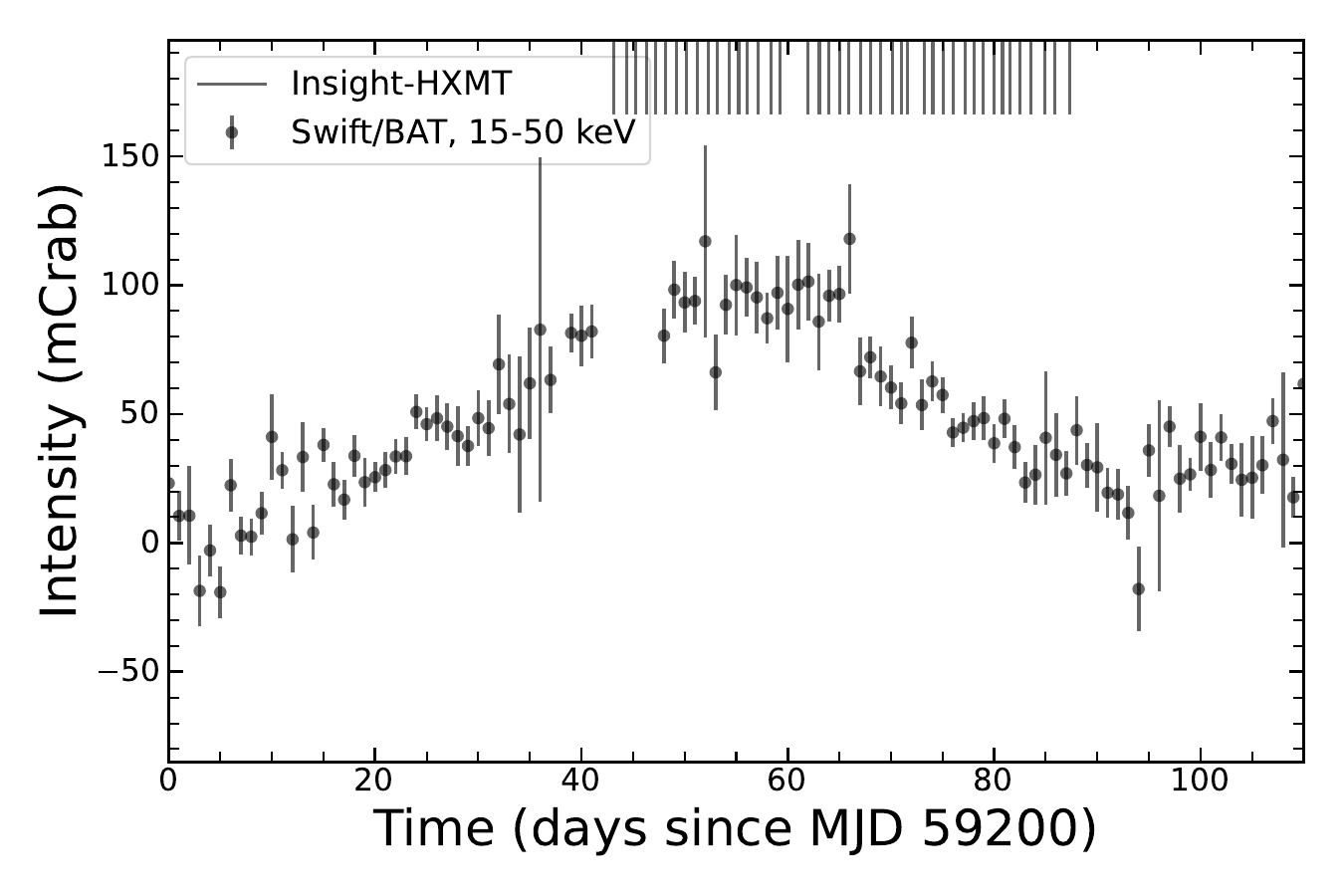}
    \caption{Light curve of 2S 1417-624 at 15-50 keV, as observed by \textit{Swift}/BAT during the 2018 and 2021 outbursts. Upper panel: Total of 29 pointing observations conducted by Insight-HXMT in 2018, as indicated by the red points with the flux in the 2-140 keV energy band. Bottom panel: Vertical black lines indicating the Insight-HXMT observations in 2021. Table \cref{tab:ObsIDs2021} provides details of the observational information.}
    \label{fig:counts}
\end{figure}

The Insight-HXMT satellite is equipped with three scientific payloads: the High Energy X-ray telescope (HE; 20–250 keV), the Medium Energy X-ray telescope (ME; 5–30 keV), and the Low Energy X-ray telescope (LE; 1–15 keV). The effective areas of these instruments are 5000 $\rm cm^2$, 952 $\rm cm^2$, and 384 $\rm cm^2$, with time resolutions of 25 $\mathrm{\mu}$s and 276 $\mathrm{\mu}$s, and 1 ms for the HE, ME, and LE telescopes, respectively. Insight-HXMT conducted a total of 29 point observations of 2S 1417-624 \citep{2020SCPMA..6349502Z} from April 13, 2018, to July 11, 2018, as indicated by the {red points in the upper panel of Fig. \cref{fig:counts}}, and 43 observations from January 19, 2021, to March 14, 2021, as {represented by the vertical black lines in the bottom panel} of Fig. \cref{fig:counts}.

For data reduction the Insight-HXMT Data Analysis Software (HXMTDAS v2.04) was employed. The tasks \textsl{he/me/legtigen} were utilised to generate a good time interval (GTI) file with pointing offset angles $<0.04^\circ$, elevation angles $>10^\circ$, and geomagnetic cut-off rigidity $>8$ GeV, while eliminating intervals associated with the South Atlantic Anomaly passage (300s before and after SAA passage). The tasks \textsl{he/me/lescreen} were used to screen the data. The arrival times of events were subsequently corrected to the Solar system barycenter using the \textit{hxbary} task. Light curves were extracted using the tasks \textit{he/me/lelcgen} with a bin size of 0.0078125~s (1/128~s).

For spectral analysis the tasks \textsl{he/me/lespecgen} were employed to generate spectra, with energy ranges of 2–10 keV for LE, 10–30 keV for ME, and 30–140 keV for HE. The tasks \textsl{he/me/lebkgmap} were used to estimate the background (\citealt{2020JHEAp..27...24L,2020JHEAp..27...44G,2020JHEAp..27...14L}). The XSPEC 12.13.1 \citep{1996ASPC..101...17A} was utilised for the spectral fitting analysis, {adopting the $\chi^2$ minimisation statistic method}. Uncertainties were estimated using the Markov chain Monte Carlo (MCMC) method with a chain length of 20,000 steps, reported at the 68\% confidence level unless otherwise noted. Given the current accuracy of instrument calibration, systematic errors of 0.5\% were adopted for the spectral analysis. In addition, we grouped the spectra from LE, ME, and HE with a minimum of 25 counts per bin.

\section{Results}\label{sec:results}
\subsection{Pulse profiles}\label{subsec:timing}

\begin{figure*}
    \centering
    \includegraphics[width=.33\textwidth]{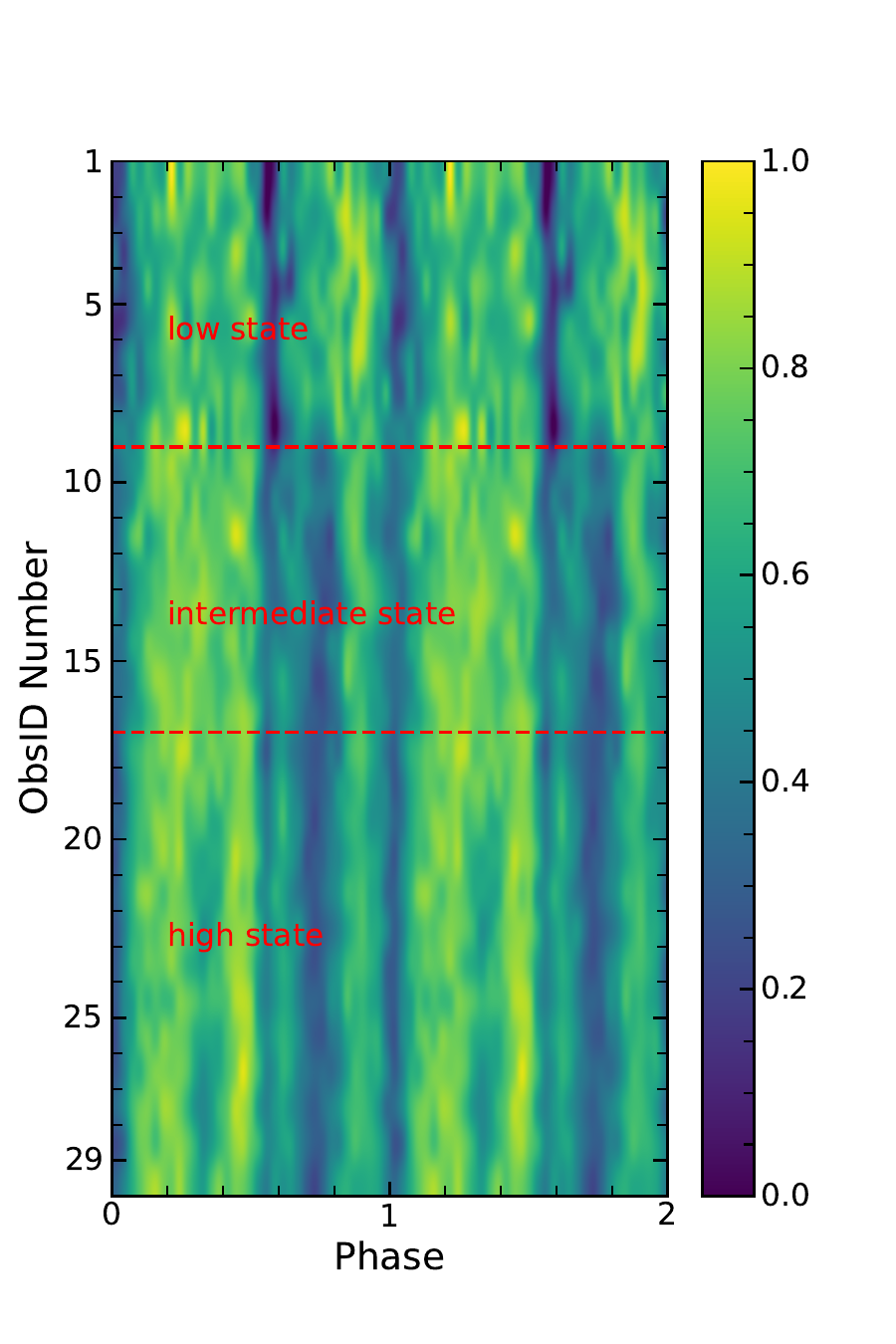}
    \includegraphics[width=.33\textwidth]{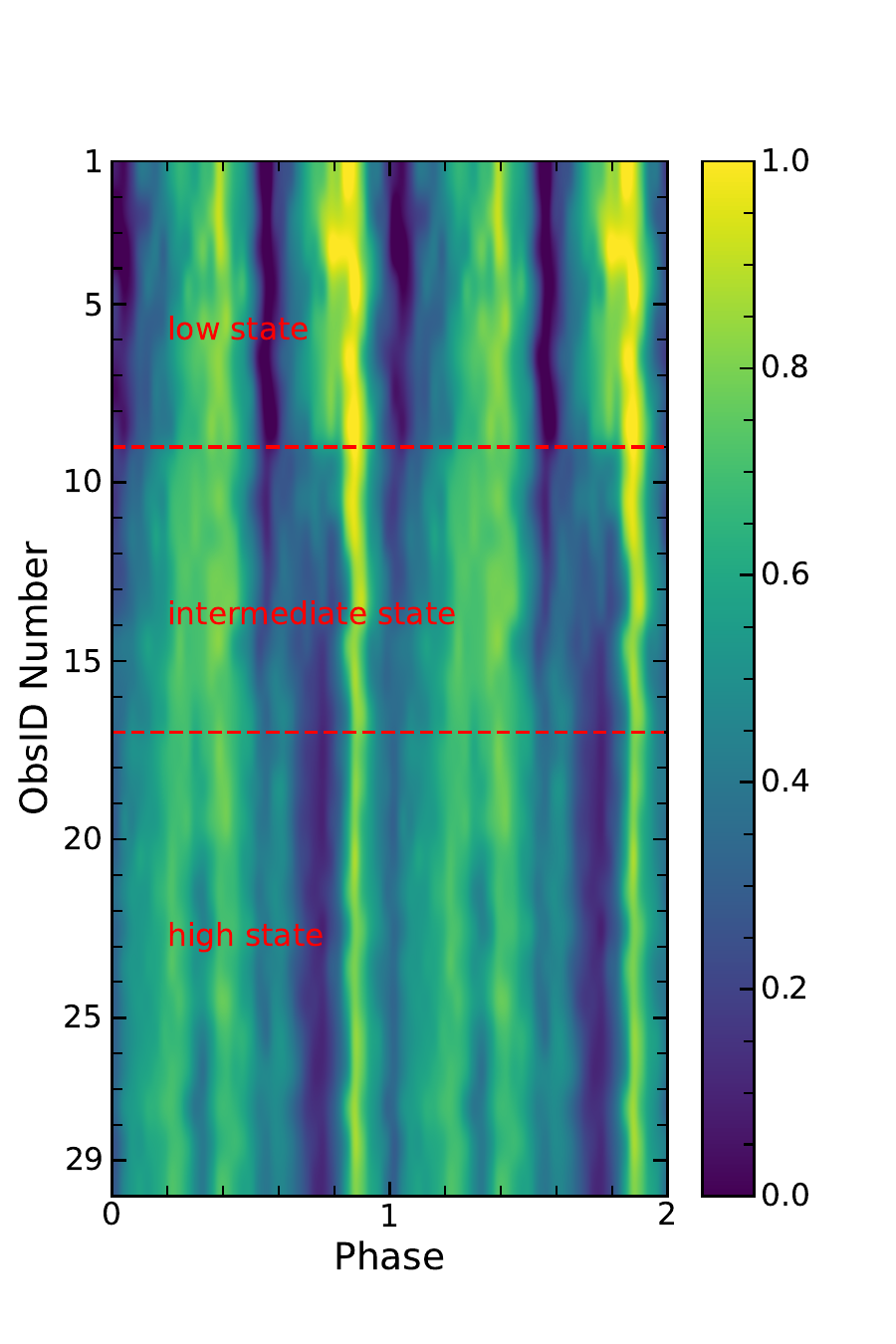}
    \includegraphics[width=.33\textwidth]{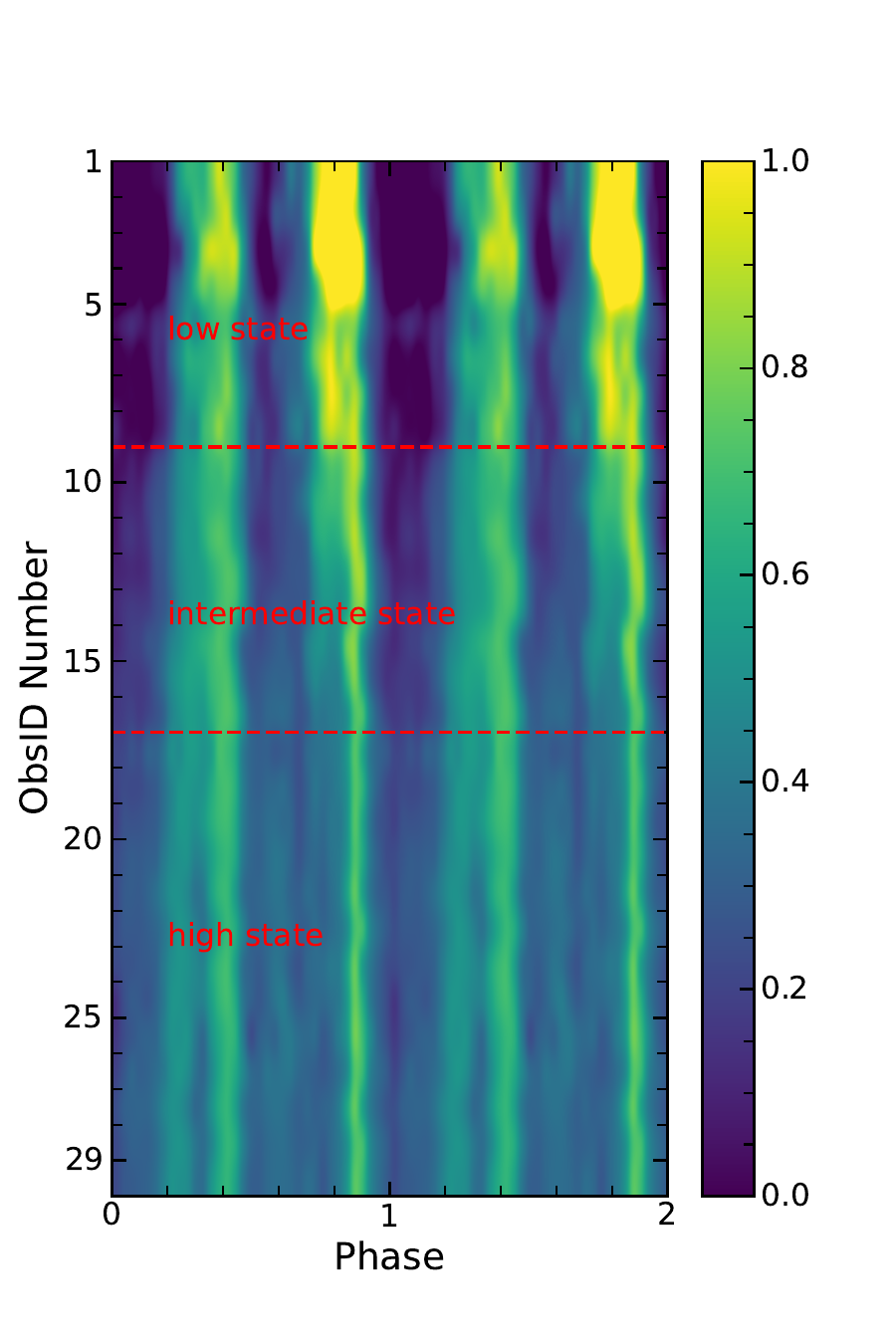}
    \caption{Pulse profile map normalised to the range of 0--1 for all Insight-HXMT observations of the 2018 outburst, sorted by ObsID number (in ascending order of flux). This figure illustrates the flux dependence of the pulse profiles in the LE (the left panel), ME (the middle panel), and HE (the right panel) energy bands. The dashed red lines indicate the two flux levels (ObsID $\sim$ 9 and 17) where the profile shape changes.}
    \label{fig:pulse}
\end{figure*}


\begin{figure*}
    \centering
    \includegraphics[width=.33\textwidth]{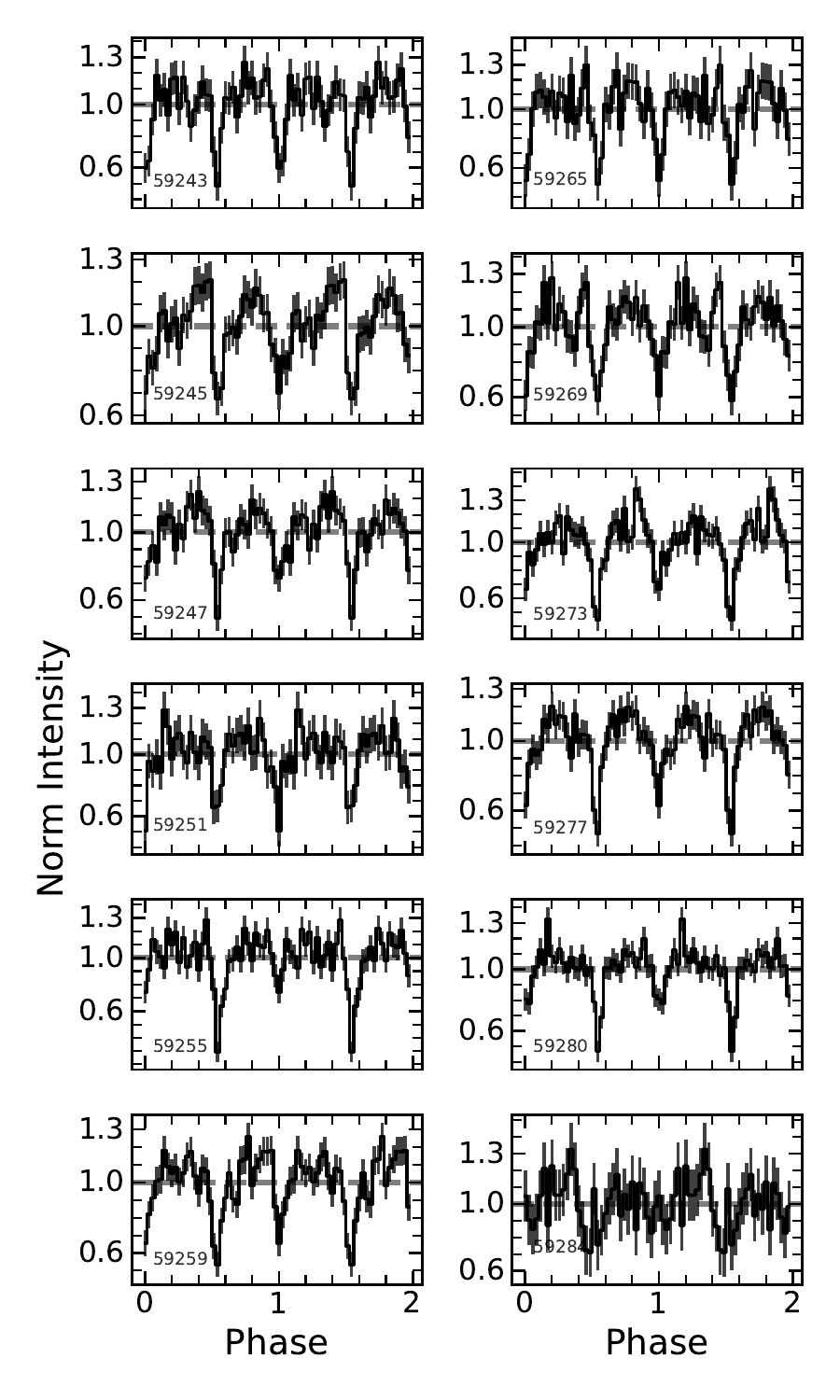}
    \includegraphics[width=.33\textwidth]{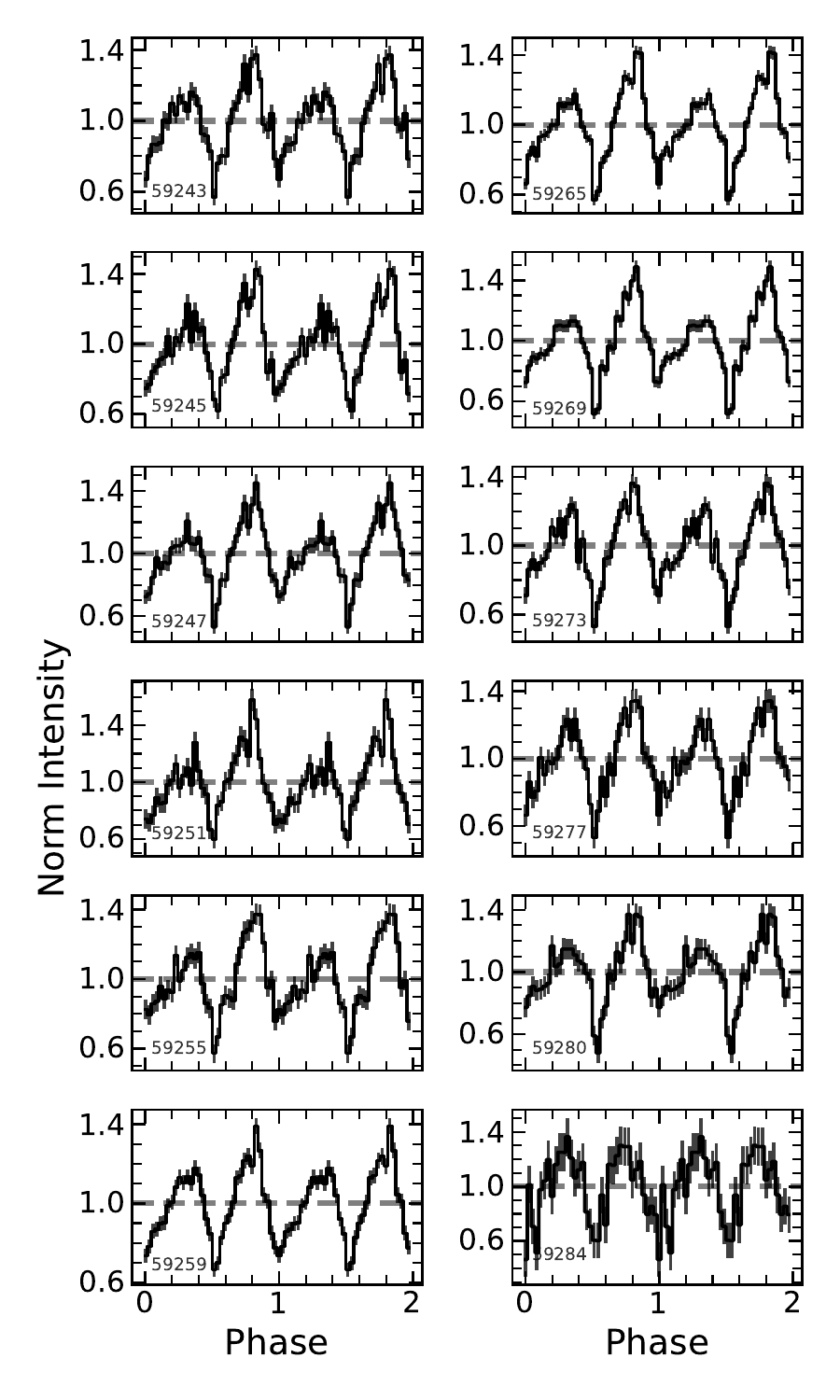}
    \includegraphics[width=.33\textwidth]{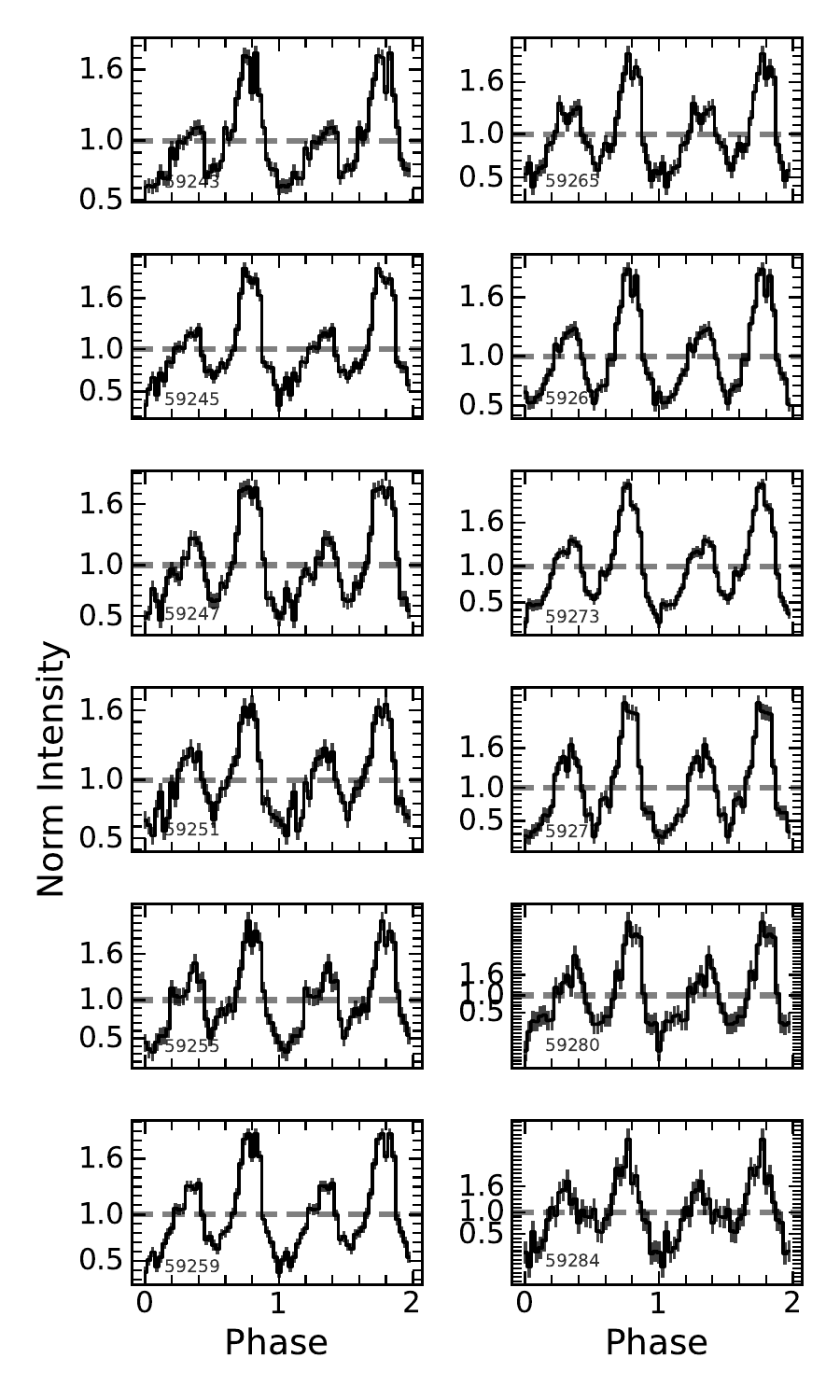}
    \caption{Pulse profiles of 2S 1417-624 in the energy range of 1-10 keV (the left panel), 10-30 keV (the middle panel), and 30–100 keV (the right panel), from Insight-HXMT/(LE/ME/HE) data at different times (in MJD, left hand of each panel) during the 2021 outburst.}
    \label{fig:pulse2021}
\end{figure*}

To find the spin period of 2S 1417-624 for each observation, we performed a pulsation search using the epoch folding technique with the \textit{efsearch} tool of HEASoft 6.32.1. The optimal pulse period was identified by locating the maximum $\chi^{2}$ value. {The uncertainty on the spin period (at the 68\% confidence level) was estimated by generating 5000 synthetic light curves, each sampled with the same cadence as the original data and constructed by drawing count rates randomly from Poisson distributions. For each simulated light curve, the pulse period was determined using epoch folding. The standard deviation of the resulting distribution of best-fit periods was adopted as the 1$\sigma$ uncertainty.} Subsequently, the pulse profiles were generated by folding each light curve at the derived best spin period. For the 2018 outburst, we plotted the pulse profiles map for the LE (1–10 keV), ME (10–30 keV), and HE (30–100 keV) energy bands and found the significant evolution in the shape of the pulse profiles as a function of ObsID number, sorted in ascending order of flux (here the fluxes estimated from the spectral fits are listed in Table \cref{tab:ObsIDs}). As shown in Fig. \cref{fig:pulse}, all pulse profiles in the LE (the left panel), ME (the middle panel), and HE (the right panel) energy bands exhibit broad double peaks at the low state below $\sim$4.1$\times 10^{-9}$ erg cm$^{-2}$ s$^{-1}$ (ObsID $\sim$ 9). At fluxes exceeding this flux threshold, i.e. in the intermediate state, the 1–10 keV pulse profile evolves from double to triple peaks, while in the 10–30 keV and 30–100 keV energy bands, the peak in the 0.8 phase becomes narrower. This different morphology of the profiles in LE compared to ME and HE bands suggests that the emission pattern is likely a combination of ‘pencil’ and ‘fan’ beams. At high state, beyond $\sim$6.4$\ \times \ 10^{-9}$ erg cm$^{-2}$ s$^{-1}$ (ObsID $\sim$ 17), another peak splits into two sub-peaks, the 1-10 keV profile evolves from triple to quadruple peaks structure, and the 10–30 keV and 30–100 keV profiles exhibit quadruple and triple peaks, respectively. This evolution, particularly the transition from double to triple peaks in the 1–10 keV band followed by a shift to quadruple peaks, is consistent with NICER observations reported by \cite{2020MNRAS.491.1851J}. They propose two transitions: the first reflecting the shift from a sub-critical to a supercritical accretion regime \citep{2018MNRAS.479.5612G}, and the second indicating the transition of the accretion disc from the gas-dominated to the radiation pressure-dominated state. Apart from this scenario, other possibilities can be considered. For instance, as \cite{2012A&A...544A.123B} predicted, in the high states, the X-ray radiation escapes through the accretion column walls of the emitting region, forming a ‘fan’ beam. At low states, the emission escapes from the top of the column, forming a ‘pencil’ beam. In intermediate states, the emission pattern may be a combination of these two cases. Therefore, the transition from double peaks into triple peaks in LE bands may indicate that the accretion of this system approaches the critical regime, suggesting that the beam pattern changes from a ‘pencil’ beam to a mixture of the ‘pencil’ and ‘fan’ beams. This is reflected in the narrowing of the pulse profile in the higher energy band of 10–100 keV. Furthermore, the evolution from a triple-peaked profile at intermediate states to a four-peaked profile at high states in the low-energy band, and from a narrowing double-peaked to a multi-peaked profile in the high-energy band may imply a full transition from the sub-critical to a supercritical regime, i.e. a change in the emission pattern from a combination of ‘pencil’ and ‘fan’ beams to a predominantly ‘fan’ beam configuration, similar to the Swift J0243.6+6124 \citep{2018ApJ...863....9W}. \cite{Liu_2024} find an anti-correlation between pulse fraction and flux in different energy bands in the range of $\sim$(1--6)$\ \times \ 10^{-9}$ erg cm$^{-2}$ s$^{-1}$, which has also been reported in recent studies \cite{2018MNRAS.479.5612G,2019MNRAS.490.2458G}. \cite{Liu_2024} also argues that the pulse fraction has a minimum around the transitional flux of $\sim$6$\times 10^{-9}$ erg cm$^{-2}$ s$^{-1}$. Above that flux threshold, it remains stable as the flux rises and has an increasing trend, supporting the interpretation that the transition from the sub-critical to the supercritical accretion regime occurs.

For the outburst in 2021, the pulse profiles showed double peaks and did not evolve with flux because the flux intensity was very low compared to the 2018 outburst. Here, we present 12 observations in approximately one orbit to show the pulse profile with time. Figure \cref{fig:pulse2021} shows the pulse profile obtained at different times in 2021 in three energy ranges. As we can see, the pulse profile has a prominent double peak. As time progresses, they do not evolve over the orbital phase and remain stable. The two peaks seem to be comparable in the LE band. As the energy increases, the second peak of the profiles becomes stronger and dominates.

\subsection{The orbital parameters}

\begin{figure}
\centering
\includegraphics[width=0.5\textwidth]{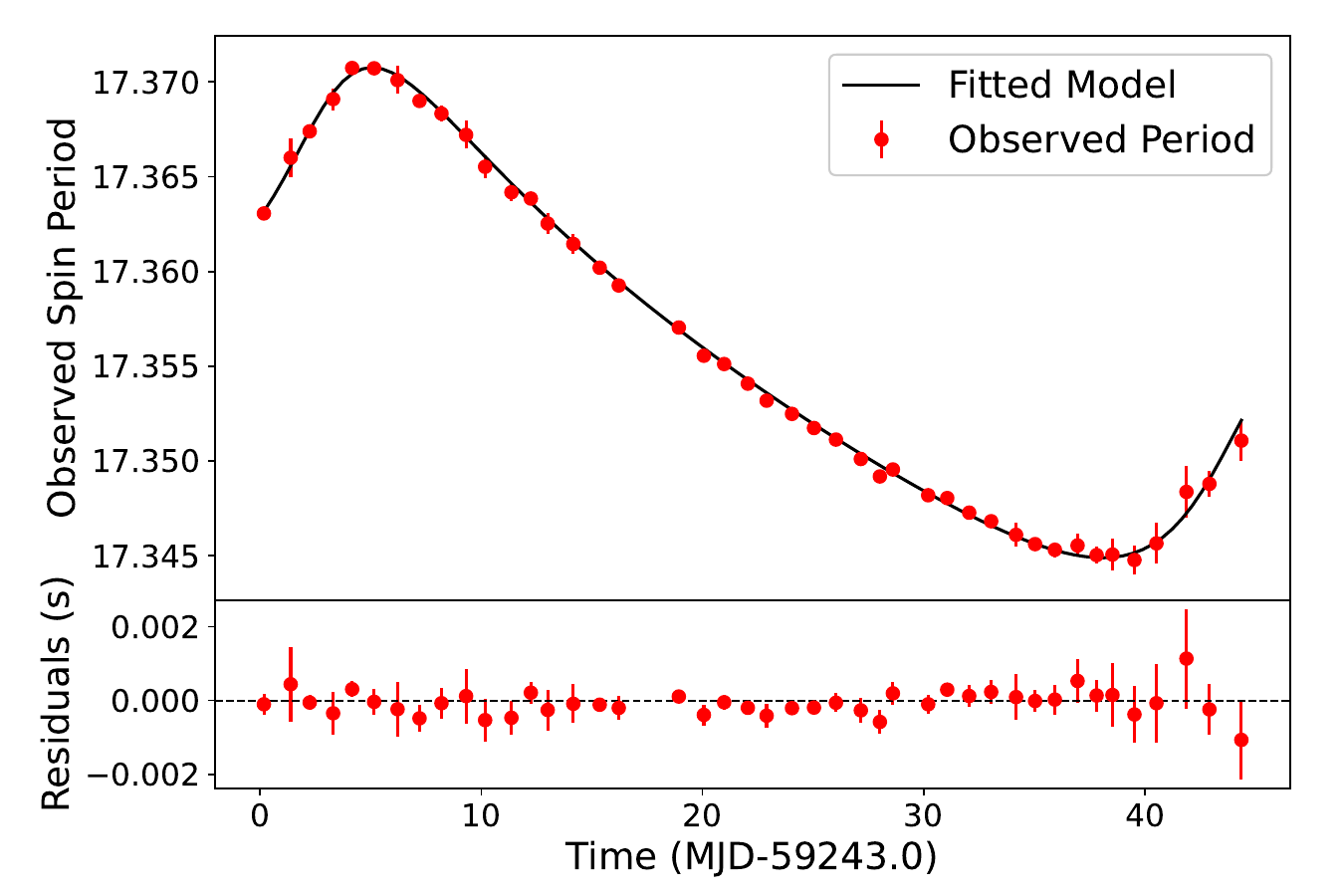}

\caption{Observed spin period values of 2S 1417-624, during the 2021 outburst based on Insight-HXMT. The solid black curve represents the best-fitting model due to the Doppler motion of the binary. Residuals are shown in the lower panel.}
\label{fig:orbit}
\end{figure}

\begin{table*}[h!]
\centering
\caption{Orbital parameters of 2S 1417--624.}
\label{tab:orbital}
\label{table:orbital_parameters}
\begin{tabular}{lcccc}
\hline
\hline
Parameter & \multicolumn{2}{c}{2021 outburst$^a$} &  1999 outburst$^b$ & 1994 outburst$^c$ \\
\hline
$P_{\text{spin}}$ (s) & $17.3638 \pm 0.0002$ & $17.3638 \pm 0.0017$ & $17.544701 \pm 0.000008$ & -- \\
$\dot{P}_{\text{spin}}$ (s~d$^{-1}$) & $(-3.60 \pm 0.12) \times 10^{-4}$ & $(-3.55 \pm 2.24) \times 10^{-4}$ & $(-3.46 \pm 0.56) \times 10^{-5}$ & -- \\
$\ddot{P}_{\text{spin}}$ (s~d$^{-2}$) & --& $(-0.2 \pm 10.0) \times 10^{-6}$ & $(-3.42 \pm 0.47) \times 10^{-6}$ & -- \\
$a_x \sin i$ (light-seconds) & $221.4 \pm 5.6 $ & $221.6 \pm 9.6 $ & $207.05 \pm 0.97$ & $188 \pm 2$ \\
$\omega$ ($^\circ$) & $303.6 \pm 1.7 $ & $303.9 \pm 11.2$ & $298.85 \pm 0.68$ & $300.3 \pm 0.6$ \\
$T_\omega$ (MJD) & $59245.74 \pm 0.17$ & $59245.74 \pm 0.31$ & $51485.059 \pm 0.057$ & $49713.62 \pm 0.05$ \\
$e$ & $0.3996 \pm 0.0099$ & $0.399 \pm 0.027$ & $0.4169 \pm 0.0033$ & $0.446 \pm 0.002$ \\
\hline
\end{tabular}
\begin{tablenotes}
\item $^a$ $P_{\text{spin}}$ at MJD 59243.0.
\item $^b$ Values from \cite{2010MNRAS.406.2663R}.
\item $^c$ Values from \cite{Finger1996}.
\end{tablenotes}
\end{table*}

In general, the measurement of the pulse arrival times can be used to estimate the orbital parameters in pulsar binaries. However, the arrival time may be affected by the pulse profile shapes. In this work, we used the measured spin period (e.g. \citealt{2010MNRAS.406.2663R}) to determine the orbital parameters. The observed pulse periods, $p_{\mathrm{obs}}$, are modified by the binary orbital motion. Due to the Doppler effect, the spin period, \(P_{\text{spin}}^{\text{obs}}(t)\), of the neutron star is given based on the following relation:
\begin{equation}
P_{\text{spin}}^{\text{obs}}(t) = \left( P_{\text{spin}}(t_0) + \dot{P}_{\text{spin}}(t - t_0) \right) 
\sqrt{\frac{1 + v_r/c}{1 - v_r/c}},
\end{equation}
{where} \(\dot{P}_{\text{spin}}\) represents the rate of change of the spin period, \(P_{\text{spin}}^{\text{obs}}\) is the spin period observed at time \(t\), and \(t_0\) denotes the start time of the outburst. The velocity component along the line of sight of the neutron star, \(v_r\), is determined by the orbital elements:
\begin{equation}
v_r = \frac{2 \pi a_x \sin i}{(1 - e^2)^{1/2} P_{\text{orb}}} 
\left( \cos(\nu + \omega) + e \cos \omega \right).
\end{equation}
In this context \(a_x \sin i\) represents the projected semi-major axis; \(P_{\text{orb}}\) is the orbital period; \(e\) is the eccentricity; and \(\omega\) is the angle of periastron of the neutron star's orbit. The true anomaly, \(\nu\), is related to the time of periastron passage, \(T_\omega\), through the eccentric anomaly, \(E\), as follows:
\begin{equation}
\tan \frac{\nu}{2} = \sqrt{\frac{1 + e}{1 - e}} \tan \frac{E}{2},
\end{equation}
\begin{equation}
E - e \sin E = \frac{2 \pi}{P_{\text{orb}}}(t - T_\omega).
\end{equation}

The data we used are from the 2021 outburst by Insight-HXMT because it covers about one orbit and has many more consecutive observations (for more details see Table \cref{tab:ObsIDs2021}). We fitted the observed spin periods with the function above. The fitting is based on the least-squares fitting. The errors on the spin period were multiplied by a constant factor to get a reduced $\chi^2$ of $\sim$1 and were weighted when the orbital parameters were fitted. The estimated error for orbital parameters was computed from the covariance matrix.

To test the performance, we first performed a fitting with all the parameters free. We found the orbital period to be 43.08 ± 0.50 d, which is consistent with the value of 42.175 d derived from \textit{Fermi}/GBM\footnote{\url{https://gammaray.msfc.nasa.gov/gbm/science/pulsars/light curves/2s1417.html}}. However, because the obtained parameters had large error bars, we fixed the orbital period, $P_{\mathrm{orb}}$, to a value of 42.175 d to obtain a good constraint. Figure \cref{fig:orbit} shows the best-fitting plots from the all observations in 2021 with Insight-HXMT; a clear sinusoidal variation is present. The fitting results, including the spin period at t0, the eccentricity, and the projected semi-major axis for the observations are listed in Table \cref{tab:orbital}. The results for the 2021 outburst are generally consistent with earlier measurements under certainties {(e.g. \cite{Finger1996} and \cite{2010MNRAS.406.2663R}), although the uncertainties in our measurements are larger. While the absolute values of the parameters deviate somewhat from those reported in previous studies, these differences remain within the quoted uncertainties and are likely due to the limited statistical quality of the data. Nevertheless, our analysis provides a new measurement of the orbital parameters based on the most recent outburst. We note that $\dot{P}_{\text{spin}}$ may significantly depend on the accretion rate. Although the 2021 outburst was relatively weak with modest accretion variability, we evaluated the impact of assuming a constant $\dot{P}_{\text{spin}}$ by including a second derivative term, $\ddot{P}_{\text{spin}}$, in the binary orbital model. We find that the fitted orbital parameters remain consistent (see Table \cref{tab:orbital}), though with notably larger uncertainties.}

\subsection{Spectral analysis}\label{subsec:spectral}

The broadband spectra of {X-ray pulsars} are typically characterised by a cut-off power law or a high-energy cut-off. In this work, we applied a common continuum model—a cut-off power law previously used for 2S 1417-624 \citep{2018MNRAS.479.5612G, 2019MNRAS.490.2458G, 2020MNRAS.491.1851J,Liu_2024}—to fit the spectra of 2S 1417-624, which is characterised by
\begin{equation}
C(E)=N \cdot E^{-\Gamma}\exp \left(\frac{-E}{E_{\text {cut }}}\right),
\end{equation}
where \( \Gamma \) is the power-law photon index and \( E_{\text{cut}} \) is the cut-off energy. The iron emission line around 6.4 keV, \( f_{\rm gauss} \) (Gaussian in XSPEC) and the photoelectric absorption were also included in our models. The energy of the iron line was fixed at 6.48 keV \citep{Liu_2024}, and the photoelectric absorption \( f(N_{\rm H}) \) (TBabs in XSPEC) component used the abundances from {\tt wilm} \citep{2000ApJ...542..914W} and the {\tt vern} cross-sections \citep{1996ApJ...465..487V}. Since the hydrogen column density could not be constrained well, we also fixed it at 1.6 $\ \times \ 10^{22}$ atoms $cm^{-2}$ according to the previous study \citep{Liu_2024}. \cite{Liu_2024} include a thermal blackbody component to describe a soft excess. We note that the contribution in our spectra is only prominent at high flux states and can be neglected at the low flux below $\sim$4.0$\ \times \ 10^{-9}$ erg cm$^{-2}$ s$^{-1}$. The CRSF around 100 keV in a single observation is present but statistically weak \citep{Liu_2024}; therefore, we ignored this cyclotron line component in our spectral analysis. The normalisation constants were allowed to vary independently between the LE, ME, and HE spectra. Consequently, the total model adopted in this analysis is:
\begin{equation}
S(E)= constant \times f({\rm N_H }) \times (C(E) + f_{\rm gauss} + bbodyrad).
\end{equation}

\begin{figure}
    \centering
    \includegraphics[width=.49\textwidth]{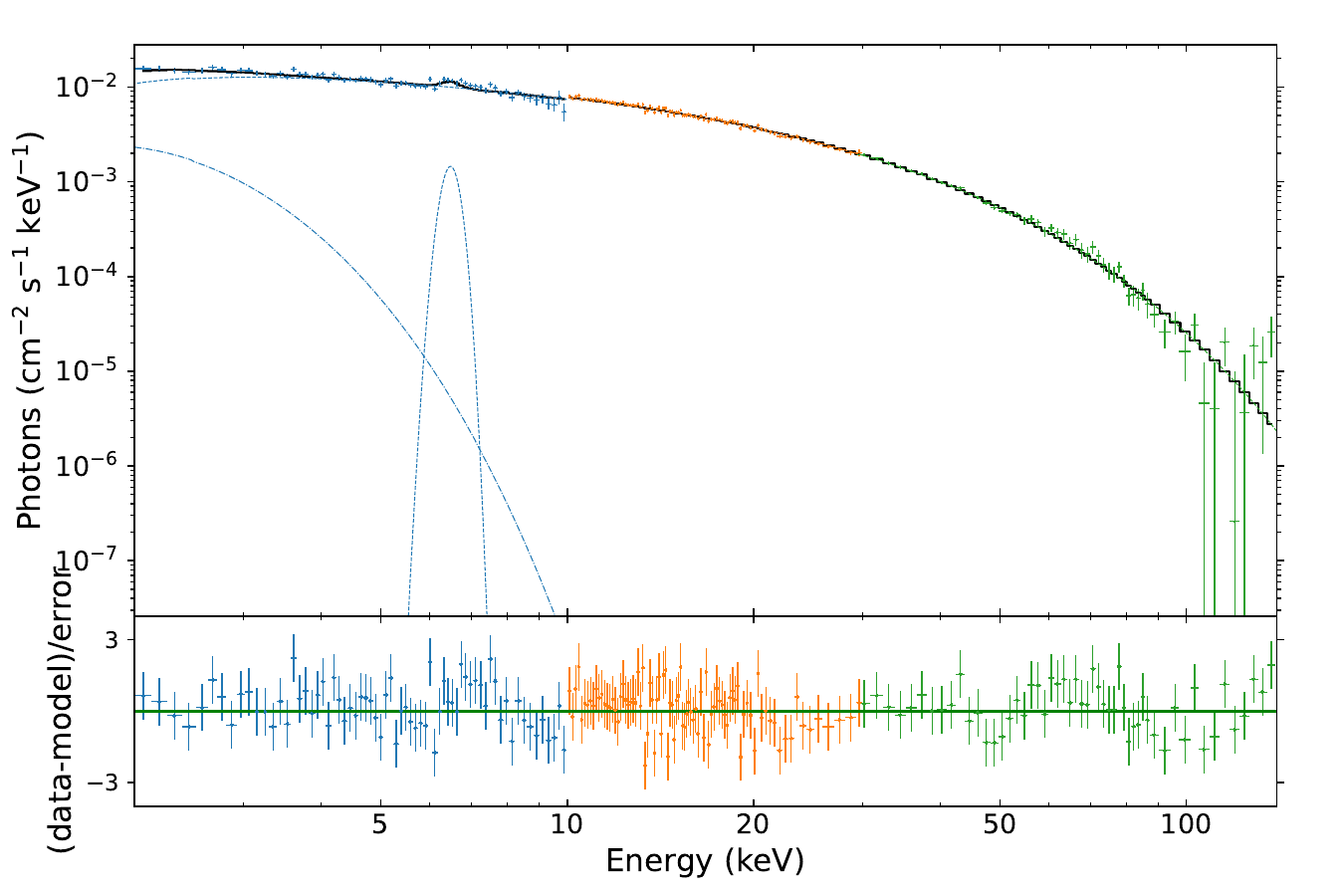}
    \caption{X-ray fitting spectrum of 2S 1417-624 over the energy range of 2–140 keV for the observation P011470900401, along with the residuals of the best-fitting model.}
    \label{fig:spectrum}
\end{figure}

The spectra were fitted by the model over the energy range of 2–10 keV for LE, 10–30 keV for ME, and 30–140 keV for HE. As an example, Fig. \cref{fig:spectrum} presents the best-fitting spectra for the observation P011470900401, along with the residuals between the data and the model with the reduced $\chi^2$ of 0.84 for 1256 d.o.f. The power-law photon index, $\Gamma$, is $0.24_{-0.03}^{+0.01}$, and the width of the iron line is around $0.20_{-0.07}^{+0.31}$ keV. The cut-off energy $E_{\rm cut}$ and the blackbody temperature kT$_{bb}$ are determined to be $17.34_{-0.39}^{+0.24}$ keV and $0.51_{-0.06}^{+0.06}$ keV, respectively. The estimated flux is measured to be $5.06_{-0.02}^{+0.04}$ $\times 10^{-9}$ erg cm$^{-2}$ s$^{-1}$. These continuum parameters are consistent with \cite{Liu_2024}. Spectral fitting was performed for all the observations, and the best-fit parameters are listed in Table \cref{tab:ObsIDs} and Table \cref{tab:ObsIDs2021} for the 2018 and 2021 outbursts, respectively.

We also tried a physical CompTT model to fit this spectrum. We find that this model nonetheless yields a poor fit with a reduced \( \chi^2 \sim 2 \). \cite{2007ApJ...654..435B} (BW) proposed a physical model for accretion-powered neutron stars based on the radiative transfer equation within the accretion column. According to their model, seed photons from bremsstrahlung, blackbody, and cyclotron emissions undergo thermal and bulk Comptonisation in the infalling plasma. This model was successfully applied to the pulsar 4U 0115+63 by \cite{2009A&A...498..825F}. Using the same approach, we attempted to fit the spectra of 2S 1417-624 for the observation mentioned above, i.e. ObsID P011470900401 (intermediate-flux state). The accretion rate, \( \dot{M} \), was fixed by the observed luminosity, assuming a distance of 13 kpc \citep{Liu_2024}. We initially obtained a good fit with the column radius, \( r_{0} \), as a free parameter. Since \( r_{0} \) depends on the \( \dot{M} \) \citep{2009A&A...498..825F}, we then fixed \( r_{0} \sim 23.0 \) m to obtain the best fit. For a typical neutron star mass and radius, we obtain an electron temperature \( T_{e} \) of $\sim$10.2 keV, with a reduced \( \chi^2 \) of 0.97 with 1354 d.o.f. However, these parameters, particularly the magnetic field B in the BW model, are insensitive and challenging to constrain. Thus, we obtain only an estimated value of $B \sim$$9.2 \times 10^{12} \, \mathrm{G}$ without associated uncertainties, consistent with the value derived from CRSF energy reported by \cite{Liu_2024}.

\begin{figure}
    \centering
    \includegraphics[width=.5\textwidth]{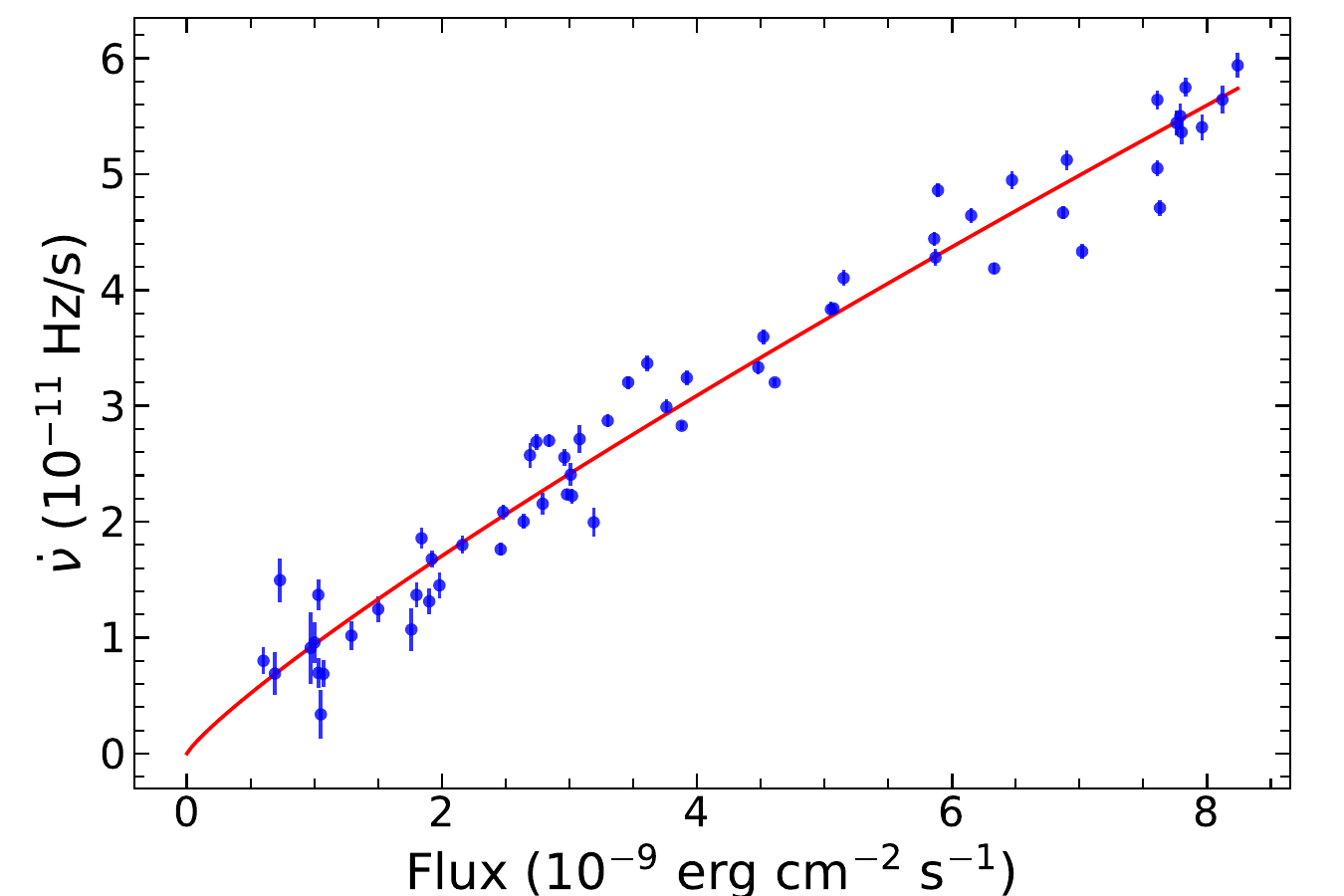}
    \caption{Spin frequency derivatives as a function of the 2–140 keV flux derived from \textit{Fermi}/GBM observations in 2018. The red line represents the best fit using the GL model.}
    \label{fig:GL}
\end{figure}

\begin{figure}
    \centering
    \includegraphics[width=.5\textwidth]{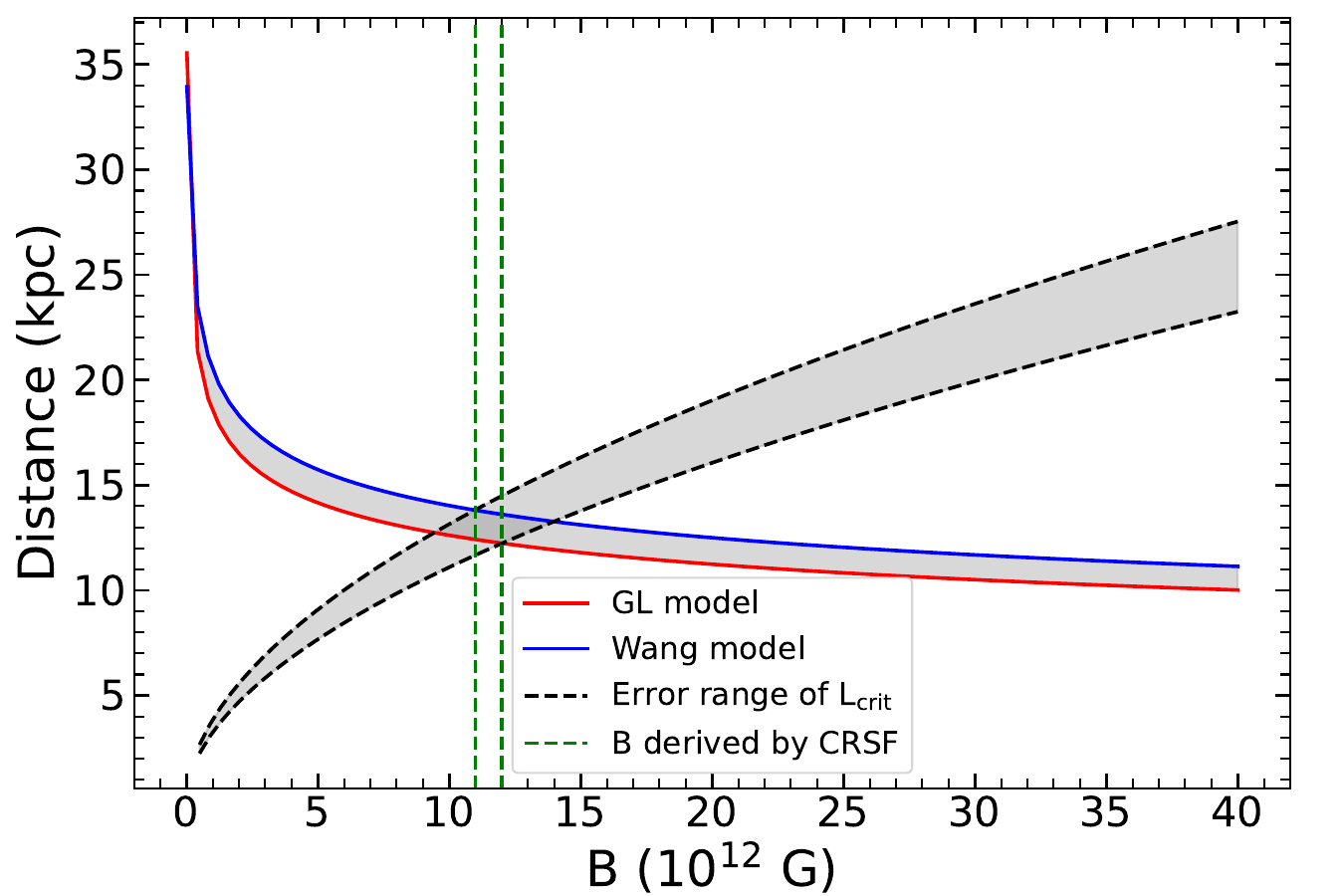}
    \caption{Relation between the distance and the magnetic field strength (D--B relation) from the GL model \citep{1979ApJ...234..296G}, Wang model \citep{1995ApJ...449L.153W}, and critical luminosity model \citep{2012A&A...544A.123B}.}
    \label{fig:DB}
\end{figure}

\section{Discussion} \label{sec:discussion}

\begin{figure*}
    \centering
    \includegraphics[width=.49\textwidth]{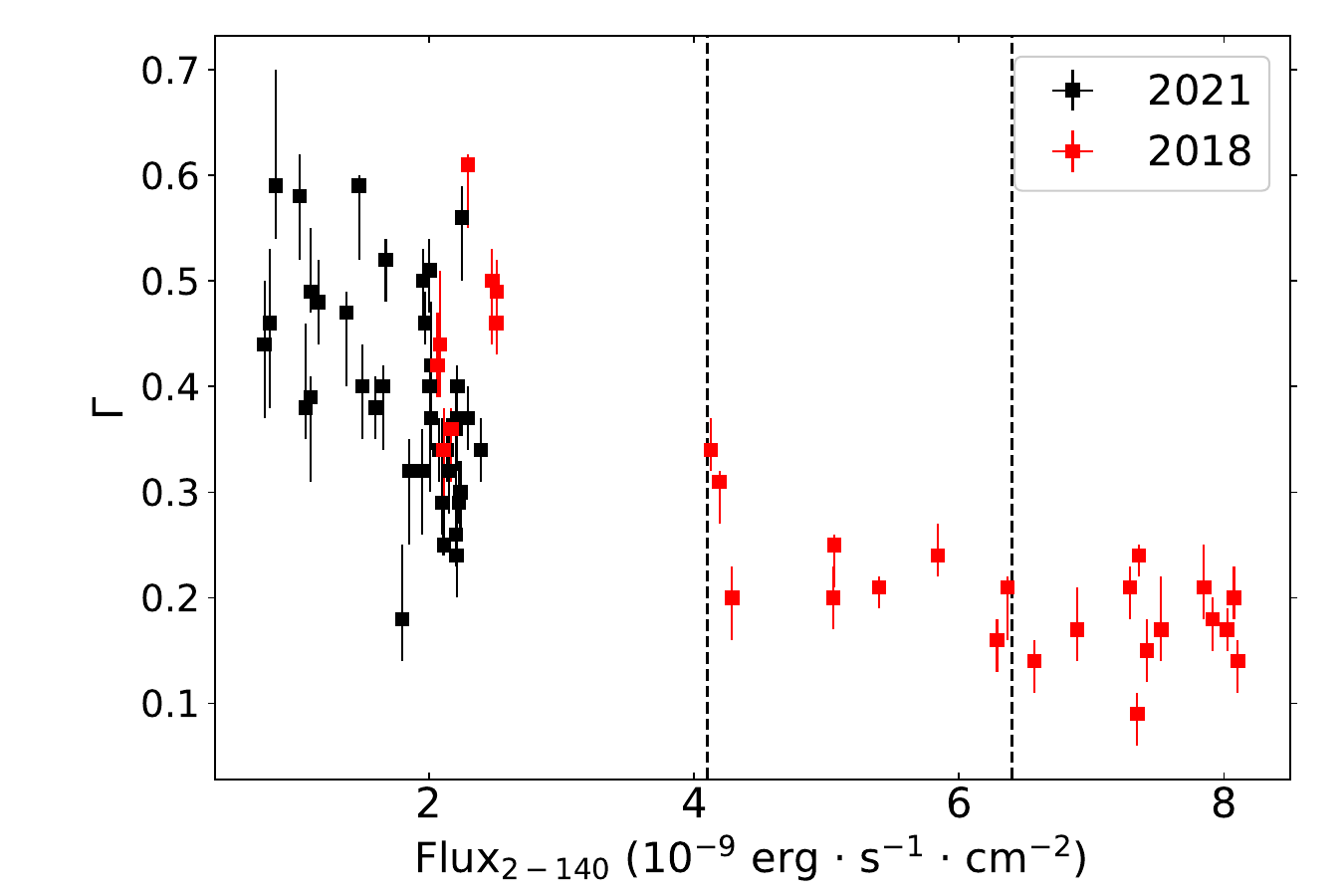}
    \includegraphics[width=.49\textwidth]{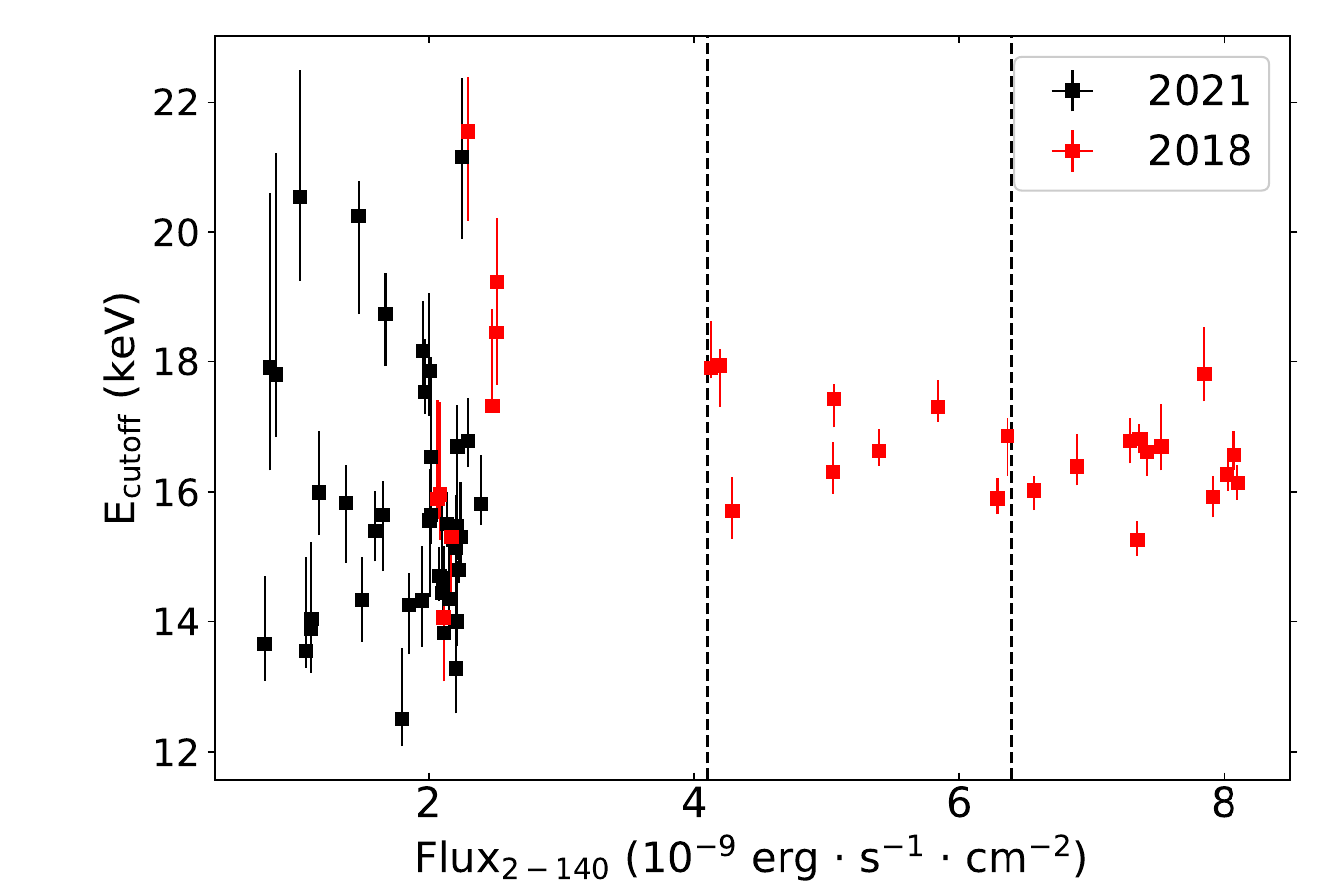}
    \caption{Spectral continuum parameters, including the photon index ($\Gamma$) and cut-off energy ($E_{\rm cutoff}$) as a function of X-ray unabsorbed flux. The dashed lines indicate the two flux levels where the profile shape changes. The red and black points indicate the observations during the 2018 and 2021 outbursts, respectively.}
    \label{fig:cont_Flux}
\end{figure*}

\begin{figure}
    \centering
    \includegraphics[width=.5\textwidth]{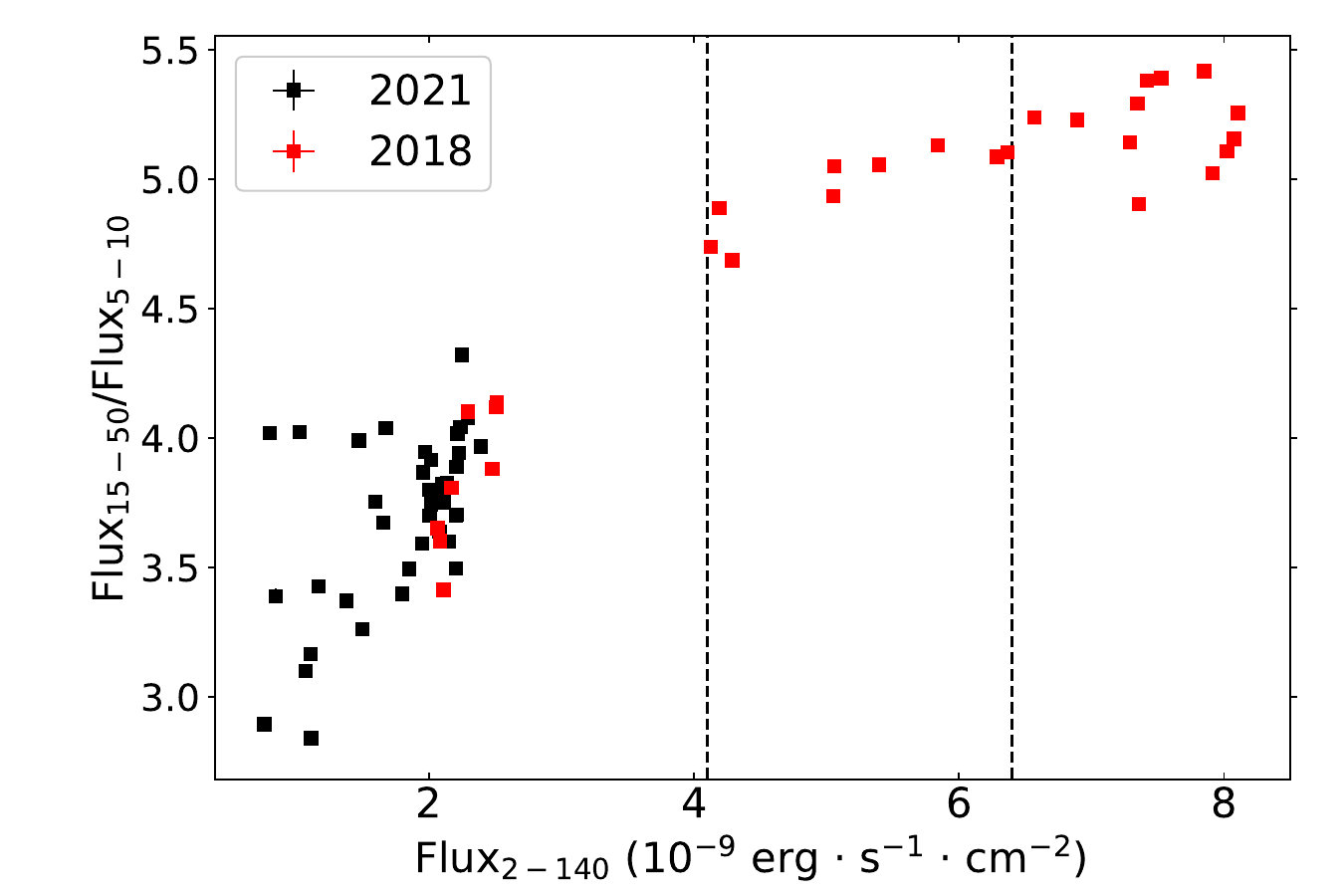}
    \caption{Hardness ratio of the fluxes in 15–50 keV and 5-10 keV ranges in 2S 1417-624 as a function of the 2-140 keV range flux. The dashed lines indicate the two flux levels where the profile shape changes. The red and black points indicate the observations during the 2018 and 2021 outbursts, respectively.}
    \label{fig:hardness}
\end{figure}

\subsection{Accretion torque models}

\cite{2020MNRAS.491.1851J} estimate the magnetic field to be approximately $7\times 10^{12}$~G and the distance to be around 20~kpc (greater than Gaia's measurement of $7.4{+3.1\atop-1.8}$~kpc) by using accretion torque models. With the accurate magnetic field strength derived from the CRSF in \citet{Liu_2024}, we aim to determine if the tension between Gaia's measurement and the accretion model estimate of the distance is reduced. Not only can this approach be used to constrain the distance, but it can also be used to verify the applicability of the accretion torque model. The accretion torque acting on the neutron star depends on its mass accretion rate and magnetic field strength. By combining the observational constraints of the spin-up rate with the accretion rate, we can eliminate the degeneracy with the distance. To further investigate the spin-up behaviour {due to the accretion torque} during the giant 2018 outburst, we analysed the spin history of the source provided by the \emph{Fermi}/GBM\footnote{\url{https://gammaray.nsstc.nasa.gov/gbm/science/pulsars/light curves/2s1417.html}} pulsar observations between MJD 58208 and MJD 58324. The frequency derivatives, $\dot{\nu} = \Delta \nu / \Delta t$, were obtained from adjacent observations, with the errors in $\dot{\nu}$ calculated using standard error propagation from the uncertainties in $\nu$. The flux corresponding to $\dot{\nu}$ was determined via linear interpolation of the \emph{Swift}/BAT count rate history, applying a conversion factor of 1.01$\times \ 10^{-7}$ erg cts$^{-1}$ \citep{Liu_2024}. The uncertainties of the flux were estimated by the count rate error. Here, we first adopted the accretion torque model of \cite{1979ApJ...234..296G} (hereafter, referred to as the GL model, assuming the magnetospheric radius, $r_0$, and Alfvén radius, $r_A$, satisfies the relation, $r_0$=0.52$r_A$), and the spin frequency derivative $\dot{\nu}$ of the pulsar is expressed as
\begin{equation}
    \dot{\nu}=5.0 \times 10^{-5} \mu_{30}^{2/7}n(\omega_{\rm{s}})R^{6/7}{M}^{-3/7}I_{45}^{-1}L_{37}^{6/7}\rm{Hz\,yr^{-1}},
\end{equation}
where $\omega_{\rm{s}}$ is the fastness parameter \citep{1977ApJ...217..578G}, and the dimensionless accretion torque, n($\omega_s$), is $\sim$1.4 for slow rotators. The moment of inertia, $I_{45} = \frac{2}{5}MR^2$, is expressed in units of $10^{45} \ \rm{g \ cm^2}$, and $\mu_{30} = \frac{1}{2}B_{12}R^3$ represents the magnetic dipole moment in units of $10^{30} \ \rm{G \ cm^3}$. Here, $B_{12}$ denotes the polar magnetic field strength at the neutron star surface in units of $10^{12}$ G; $M$ is the mass of the neutron star in solar mass units ($M_{\odot}$); $R$ is the neutron star radius in units of $10^6$ cm; and $L_{37}$ is the accretion luminosity in units of $10^{37}\ \rm{erg \ s^{-1}}$. As shown in Fig. \cref{fig:GL}, the spin frequency derivative, $\dot{\nu}$, exhibits a positive correlation with the flux. The red line in the figure represents the best fit to the data using the GL model, which reveals the following relationship between the distance $D$ and $B_{12}$: 
\begin{equation}\label{eq:11}
    D\approx (18.52 \pm 0.12) \times B_{12}^{-1/6} \ (\rm{kpc}).
\end{equation}
With this relationship, the distance can be estimated as $\sim$$12$ kpc for the magnetic field strength of $\sim$$ (1.1-1.2) \times 10^{13}$ G \citep{Liu_2024} when the gravitational redshift z=0.3 is considered. Based on another accretion model of \cite{1995ApJ...449L.153W} (the Wang model, also presented in Fig.~\cref{fig:DB}), in which $n(\omega_{\rm{s}}) \approx 7/6$ was adopted for a slow rotator, the results are different: the distance is a factor of 1.1 larger than that inferred from the GL model, i.e. $\sim$14 kpc. We note that the inferred distance of $\sim$12-14 kpc from the above accretion models is less than the $\sim$20 kpc found by \cite{2020MNRAS.491.1851J}. It is greater than Gaia’s estimation of $7.4{+3.1\atop-1.8}$~kpc but still consistent (within 2$\sigma$). {Alternatively, a quadrupole magnetic field configuration may need to be included in the torque models to explain the discrepancy in distance estimates, as suggested by \cite{2020MNRAS.491.1851J}.}

\subsection{Critical luminosity and CRSFs}

We can also estimate the distance by constraining the critical luminosity from pulse profile transitions. The pulse profiles transitioning from the sub-critical to supercritical regime is found to be at a critical level of $\sim$\( 6.4 \times 10^{-9} \, \mathrm{erg} \, \mathrm{cm}^{-2} \, \mathrm{s}^{-1} \), as depicted in Fig. \cref{fig:pulse}. This transition is associated with a switch in the emission pattern from 'pencil' to 'fan' beams. According to \citet{2012A&A...544A.123B}, we can derive the 2–140\,keV critical luminosity using the magnetic field \( B_{12} \) as
\begin{equation}
\begin{aligned}
L_{\text {crit}}\simeq 1.49 \times 10^{37} \ \mathrm{erg} \ \mathrm{s}^{-1} \times B_{12}^{16 / 15},
\end{aligned}
\end{equation}
for typical neutron star parameters: \( M_{\text{NS}} = 1.4 \, M_{\odot} \) and \( R_{\text{NS}} = 10 \, \mathrm{km} \). Although the transition seen in the pulse profiles is clear, there is still some uncertainty in determining exactly at which flux level it occurs. Considering that the transition might occur between adjacent observations, we adopted a wide critical flux value of \( (5.0-7.0) \times (1+z)^2 \times 10^{-9} \, \mathrm{erg} \, \mathrm{cm}^{-2} \, \mathrm{s}^{-1} \) in the neutron star rest frame, accounting for the gravitational redshift z = 0.3. Therefore, the relationship between distance \( D \) and \( B_{12} \) takes the following form:
\begin{equation}\label{eq:12}
    D\approx (3.25-3.85)\times B_{12}^{8/15} \ (\rm{kpc}).
\end{equation}
The distance of 2S 1417-624 inferred from the critical luminosity model is about 12-15 kpc using the magnetic field strength \(B_{12}\) derived from the CRSF energy \citep{Liu_2024}, which is consistent with the results from the accretion models. So, based on the GL, the Wang, and the critical luminosity models, we obtain a self-consistent distance range of approximately 12.0-15.0~kpc for the source 2S 1417-624 within the same magnetic field. Our study provides new distance measurements and a deeper understanding of both the torque models and the critical luminosity model in the strong magnetic field of 2S 1417-624.

The detection of CRSFs provides a direct estimation of the magnetic field strength \(B_{12}\) \citep{Liu_2024}. There are two theoretical models for the expected \(L_{\text{crit}}\), predicted by \cite{2012A&A...544A.123B} and \cite{2015MNRAS.447.1847M} for a corresponding \(B_{12}\). To assess whether the observed critical luminosity inferred from the pulse profile transition is consistent with the theoretical models, we derived the X-ray luminosity based on the distance. We assumed that the observed critical flux is approximately within the wide range of \((5-7) \times 10^{-9} \, \mathrm{erg \, cm^{-2} \, s^{-1}}\). As discussed above in the framework of \cite{2012A&A...544A.123B}, we neglect the effect of gravitational redshift here, as the models require a high critical luminosity (\(\sim 1.5 \times 10^{38} \, \mathrm{erg \, s^{-1}}\)). Adopting a distance of 13 kpc from the accretion torque models, the corresponding critical luminosity, \(L_{\text{crit}}\), is estimated to be \((1.0-1.4) \times 10^{38} \, \mathrm{erg \, s^{-1}}\), which aligns with the theoretical \(L_{\text{crit}}\) predicted by \cite{2012A&A...544A.123B}. According to \cite{2015MNRAS.447.1847M}, the critical luminosity is {likely around \(5 \times 10^{37} \, \mathrm{erg \, s^{-1}}\) at $B_{12} \sim 10^{13}$ G, based on their plot} for \(l_0/l = 1.0\) and pure X-mode polarisation, which is significantly lower than the value derived by \cite{2012A&A...544A.123B}. Assuming Gaia's distance of 7.4 kpc, the \(L_{\text{crit}}\) is found to be \((3.0-5.0) \times 10^{37} \, \mathrm{erg \, s^{-1}}\), close to the theoretical value estimated by \cite{2015MNRAS.447.1847M}. In addition, as predicted by \cite{2012A&A...544A.123B}, in the supercritical case—above a certain critical luminosity \(L_{\rm crit}\)—the radiation-dominated shock decelerates the falling material, forming the accretion column \citep{1976MNRAS.175..395B}. Morever, the emission height in the column increases with luminosity. In the sub-critical case (below \(L_{\rm crit}\)), the radiation pressure can be neglected, and the Coulomb interaction decelerates the falling plasma, resulting in hot spots on the surface of the neutron star; consequently, the height of the line-forming region reduces with luminosity. An anti-correlation between CRSF energy and luminosity is expected at high luminosity, while a positive correlation is anticipated at low luminosity. Therefore, higher-quality observational data on the dependence of the cyclotron line on luminosity will be crucial to verify the theoretical critical luminosity in the future.

\subsection{Variability of continuum parameters versus flux}

Here, we focus on the variability of the continuum parameters with respect to the unabsorbed flux in the 2–140 keV energy range. The power-law photon index and the cut-off energy show clear variations, as depicted in Fig. \cref{fig:cont_Flux}. Varying from 0.1 to 0.6, $\Gamma$ exhibits a negative correlation with the flux, indicating that the spectra harden as the flux increases. Beyond a flux of approximately $4.1 \times 10^{-9}$ erg cm$^{-2}$ s$^{-1}$ (the left dashed line), the photon index, $\Gamma$, continues to decrease. However, at flux levels above $6.4 \times 10^{-9}$ erg cm$^{-2}$ s$^{-1}$ (the right dashed line), the photon index tends to stabilise and cluster, and then shows an increasing trend. However, the cut-off energy, ranging from 14 to 21 keV, initially exhibits dramatic variations with flux before stabilising, and then displays a positive trend.

The results reveal significant spectral changes in 2S 1417-624 during the 2018 and 2021 outbursts. At low flux states, the photon index is negatively correlated with flux at levels below $\sim$6$\times 10^{-9}$ erg cm$^{-2}$ s$^{-1}$, indicating that the spectra become harder as the flux increases. The anti-correlation between the photon index and the luminosity has also been observed in other pulsars such as 4U 0115 + 63, 1A 0535 + 262 \cite{2013A&A...551A...1R}, and Swift J0243.6 + 6214 \cite{2020ApJ...902...18K}. Beyond this flux threshold, the photon index stabilises around 0.2, and the correlation reverses, with the spectra softening--consistent with the findings of \cite{2018MNRAS.479.5612G}. Meanwhile, the cut-off energy, which serves as a signature for the electron temperature, initially exhibits large fluctuation with increasing flux and then stabilises. This indicates a significant spectral change corresponding to the second pulse profile transition at $\sim$6.4$\times 10^{-9}$ erg cm$^{-2}$ s$^{-1}$. As predicted by \cite{2012A&A...544A.123B}, in the subcritical regime, the emission height of the accretion column, located a few kilometres above the neutron star surface, decreases as luminosity increases. This leads to an increase in optical depth and, consequently, a harder spectrum, as reflected in the changes in the photon index in the flux range of $\sim$(1--6)$\times 10^{-9}$ erg cm$^{-2}$ s$^{-1}$. However, at higher flux levels, above $6 \times 10^{-9}$ erg cm$^{-2}$ s$^{-1}$, both $\Gamma$ and $E_{\rm cut}$ stabilise, consistent with the expected softening of the spectrum in the supercritical state. So, the clustering of $\Gamma$ suggests the onset of a supercritical accretion regime. Furthermore, the anti-correlation between the pulse fraction and flux, as reported by \cite{Liu_2024}, changes above the flux of $\sim$6$\times 10^{-9}$ erg cm$^{-2}$ s$^{-1}$, which also indicates a transition in the accretion regime in 2S 1417-624. Figure \cref{fig:hardness} shows the hardness ratio of the fluxes in the 15-50 keV and 5-10 keV ranges as a function of the 2-140 keV flux. The hardness ratio shows a positive correlation with the flux, reaching a maximum around $\sim$6$ \times 10^{-9}$ erg cm$^{-2}$ s$^{-1}$, after which it stabilises. \cite{2015MNRAS.452.1601P} investigated the spectral hardness ratio in several transient X-ray pulsars. They find that the hardness increases with flux at low luminosities and then saturates or even slightly decreases above the critical luminosity, consistent with our results. Therefore, the variability of the spectral parameters and the hardness ratio strongly supports the hypothesis of a transition between sub-critical and supercritical accretion regimes in the pulsar 2S 1417-624.

\section{Conclusion} \label{sec:summary}
In summary, we present a detailed timing and spectral analysis of 2S 1417-624, based on observations conducted by Insight-HXMT during the outbursts in 2018 and 2021. We confirm the changes in pulse profiles in low-energy bands (1–10 keV), transitioning from a double-peaked profile at low flux states to a triple-peaked profile at $\sim$4.1$\ \times \ 10^{-9}$ erg cm$^{-2}$ s$^{-1}$, and eventually evolving into a four-peaked shape beyond the flux level of $\sim$6.4$\ \times \ 10^{-9}$ erg cm$^{-2}$ s$^{-1}$. Moreover, we identify similar changes in pulse profiles in high-energy bands (10–30 keV and 30–100 keV) around the two transition flux levels. The double-peaked pulse profiles in the 10–30 keV and 30–100 keV bands become narrower with an increase in flux around the first transition level of $\sim$4.1 $\times \ 10^{-9} \, \text{erg} \, \text{cm}^{-2} \, \text{s}^{-1}$ and subsequently exhibit four-peaked and triple-peaked profiles at higher states ($\sim$6.4 $\times \ 10^{-9} \, \text{erg} \, \text{cm}^{-2} \, \text{s}^{-1}$), respectively. Additionally, we derive a self-consistent distance of $\sim$12.0-15.0$ \, \text{kpc}$, by applying the torque models and critical luminosity model based on the magnetic field strength inferred from the CRSF. Using the observed transition flux based on Gaia's distance of 7.4 kpc and the inferred distance of 13 kpc, the estimated critical luminosity are \((3.0-5.0) \times 10^{37} \, \mathrm{erg \, s^{-1}}\) and \((1.0-1.4) \times 10^{38} \, \mathrm{erg \, s^{-1}}\), respectively. The spectral parameters and hardness ratio also exhibit significant changes around $6 \times 10^{-9}$ erg cm$^{-2}$ s$^{-1}$, strongly supporting the viewpoint of transitions between accretion regimes from sub-critical to supercritical. 
\begin{acknowledgements}
This work is supported by the NSFC (12133007, 12173103 and 12261141691) and the National Key Research and Development Program of China (Grant No. 2021YFA0718503). The authors thank the support from the Sino-German (CSC-DAAD) Postdoc Scholarship Program, 2023 (57678375). LD acknowledges funding from DFG - Projektnummer 549824807. This work has made use of data from the Insight-HXMT mission, a project funded by China National Space Administration (CNSA) and the Chinese Academy of Sciences (CAS).
\end{acknowledgements}


\bibliographystyle{aa}
\bibliography{refer}

\begin{thebibliography}{35}
\expandafter\ifx\csname natexlab\endcsname\relax\def\natexlab#1{#1}\fi

\bibitem[{{Apparao} {et~al.}(1980){Apparao}, {Naranan}, {Kelley}, \& {Bradt}}]{Apparao1980}
{Apparao}, K.~M.~V., {Naranan}, S., {Kelley}, R.~L., \& {Bradt}, H.~V. 1980, \aap, 89, 249

\bibitem[{{Arnaud}(1996)}]{1996ASPC..101...17A}
{Arnaud}, K.~A. 1996, in Astronomical Society of the Pacific Conference Series, Vol. 101, Astronomical Data Analysis Software and Systems V, ed. G.~H. {Jacoby} \& J.~{Barnes}, 17

\bibitem[{{Bailer-Jones} {et~al.}(2021){Bailer-Jones}, {Rybizki}, {Fouesneau}, {Demleitner}, \& {Andrae}}]{2021yCat.1352....0B}
{Bailer-Jones}, C.~A.~L., {Rybizki}, J., {Fouesneau}, M., {Demleitner}, M., \& {Andrae}, R. 2021, {VizieR Online Data Catalog: Distances to 1.47 billion stars in Gaia EDR3 (Bailer-Jones+, 2021)}, VizieR On-line Data Catalog: I/352. Originally published in: 2021AJ....161..147B

\bibitem[{{Bailer-Jones} {et~al.}(2018){Bailer-Jones}, {Rybizki}, {Fouesneau}, {Mantelet}, \& {Andrae}}]{Bailer2018}
{Bailer-Jones}, C.~A.~L., {Rybizki}, J., {Fouesneau}, M., {Mantelet}, G., \& {Andrae}, R. 2018, \aj, 156, 58

\bibitem[{{Basko} \& {Sunyaev}(1976)}]{1976MNRAS.175..395B}
{Basko}, M.~M. \& {Sunyaev}, R.~A. 1976, \mnras, 175, 395

\bibitem[{{Becker} {et~al.}(2012){Becker}, {Klochkov}, {Sch{\"o}nherr}, {Nishimura}, {Ferrigno}, {Caballero}, {Kretschmar}, {Wolff}, {Wilms}, \& {Staubert}}]{2012A&A...544A.123B}
{Becker}, P.~A., {Klochkov}, D., {Sch{\"o}nherr}, G., {et~al.} 2012, \aap, 544, A123

\bibitem[{{Becker} \& {Wolff}(2007)}]{2007ApJ...654..435B}
{Becker}, P.~A. \& {Wolff}, M.~T. 2007, \apj, 654, 435

\bibitem[{{Ferrigno} {et~al.}(2009){Ferrigno}, {Becker}, {Segreto}, {Mineo}, \& {Santangelo}}]{2009A&A...498..825F}
{Ferrigno}, C., {Becker}, P.~A., {Segreto}, A., {Mineo}, T., \& {Santangelo}, A. 2009, \aap, 498, 825

\bibitem[{{Finger} {et~al.}(1996){Finger}, {Wilson}, \& {Chakrabarty}}]{Finger1996}
{Finger}, M.~H., {Wilson}, R.~B., \& {Chakrabarty}, D. 1996, \aaps, 120, 209

\bibitem[{{Ghosh} \& {Lamb}(1979)}]{1979ApJ...234..296G}
{Ghosh}, P. \& {Lamb}, F.~K. 1979, \apj, 234, 296

\bibitem[{{Ghosh} {et~al.}(1977){Ghosh}, {Lamb}, \& {Pethick}}]{1977ApJ...217..578G}
{Ghosh}, P., {Lamb}, F.~K., \& {Pethick}, C.~J. 1977, \apj, 217, 578

\bibitem[{{Grindlay} {et~al.}(1984){Grindlay}, {Petro}, \& {McClintock}}]{1984ApJ...276..621G}
{Grindlay}, J.~E., {Petro}, L.~D., \& {McClintock}, J.~E. 1984, \apj, 276, 621

\bibitem[{{Guo} {et~al.}(2020){Guo}, {Liao}, {Zhang}, {Zhang}, {Tan}, {Song}, {Lu}, {Cao}, {Chang}, {Chen}, {Du}, {Ge}, {Gu}, {Jiang}, {Jin}, {Li}, {Li}, {Li}, {Liu}, {Liu}, {Lu}, {Luo}, {Meng}, {Sun}, {Yang}, {Yang}, {You}, {Zhang}, {Zhao}, \& {Zhang}}]{2020JHEAp..27...44G}
{Guo}, C.-C., {Liao}, J.-Y., {Zhang}, S., {et~al.} 2020, Journal of High Energy Astrophysics, 27, 44

\bibitem[{{Gupta} {et~al.}(2019){Gupta}, {Naik}, \& {Jaisawal}}]{2019MNRAS.490.2458G}
{Gupta}, S., {Naik}, S., \& {Jaisawal}, G.~K. 2019, \mnras, 490, 2458

\bibitem[{{Gupta} {et~al.}(2018){Gupta}, {Naik}, {Jaisawal}, \& {Epili}}]{2018MNRAS.479.5612G}
{Gupta}, S., {Naik}, S., {Jaisawal}, G.~K., \& {Epili}, P.~R. 2018, \mnras, 479, 5612

\bibitem[{{{\.I}nam} {et~al.}(2004){{\.I}nam}, {Baykal}, {Matthew Scott}, {Finger}, \& {Swank}}]{Inam2004}
{{\.I}nam}, S.~{\c C}., {Baykal}, A., {Matthew Scott}, D., {Finger}, M., \& {Swank}, J. 2004, \mnras, 349, 173

\bibitem[{{Ji} {et~al.}(2020){Ji}, {Doroshenko}, {Santangelo}, {G{\"u}ng{\"o}r}, {Zhang}, {Ducci}, {Zhang}, {Ge}, {Qu}, {Chen}, {Bu}, {Cai}, {Cao}, {Chang}, {Chen}, {Chen}, {Chen}, {Chen}, {Chen}, {Cui}, {Cui}, {Deng}, {Dong}, {Du}, {Fu}, {Gao}, {Gao}, {Gao}, {Gu}, {Guan}, {Guo}, {Han}, {Huang}, {Huo}, {Jia}, {Jiang}, {Jiang}, {Jin}, {Kong}, {Li}, {Li}, {Li}, {Li}, {Li}, {Li}, {Li}, {Li}, {Li}, {Li}, {Li}, {Liang}, {Liao}, {Liu}, {Liu}, {Liu}, {Liu}, {Liu}, {Lu}, {Lu}, {Lu}, {Luo}, {Luo}, {Ma}, {Meng}, {Nang}, {Nie}, {Ou}, {Sai}, {Song}, {Song}, {Sun}, {Tan}, {Tao}, {Tuo}, {Wang}, {Wang}, {Wang}, {Wang}, {Wang}, {Wen}, {Wu}, {Wu}, {Wu}, {Xiao}, {Xiao}, {Xiong}, {Xu}, {Yang}, {Yang}, {Yang}, {Yang}, {Yi}, {Yin}, {You}, {Zhang}, {Zhang}, {Zhang}, {Zhang}, {Zhang}, {Zhang}, {Zhang}, {Zhang}, {Zhang}, {Zhang}, {Zhang}, {Zhang}, {Zhang}, {Zhang}, {Zhang}, {Zhao}, {Zhao}, {Zheng}, {Zhou}, {Zhou}, {Zhu}, \& {Zhu}}]{2020MNRAS.491.1851J}
{Ji}, L., {Doroshenko}, V., {Santangelo}, A., {et~al.} 2020, \mnras, 491, 1851

\bibitem[{{Kelley} {et~al.}(1981){Kelley}, {Apparao}, {Doxsey}, {Jernigan}, {Naranan}, \& {Rappaport}}]{1981ApJ...243..251K}
{Kelley}, R.~L., {Apparao}, K.~M.~V., {Doxsey}, R.~E., {et~al.} 1981, \apj, 243, 251

\bibitem[{{Kong} {et~al.}(2020){Kong}, {Zhang}, {Chen}, {Zhang}, {Ji}, {Doroshenko}, {Wang}, {Tao}, {Ge}, {Liu}, {Song}, {Lu}, {Qu}, {Li}, {Xu}, {Cao}, {Chen}, {Bu}, {Cai}, {Chang}, {Chen}, {Chen}, {Chen}, {Chen}, {Cui}, {Cui}, {Deng}, {Dong}, {Du}, {Fu}, {Gao}, {Gao}, {Gao}, {Gu}, {Guan}, {Guo}, {Han}, {Huang}, {Huo}, {Jia}, {Jiang}, {Jiang}, {Jin}, {Jin}, {Li}, {Li}, {Li}, {Li}, {Li}, {Li}, {Li}, {Li}, {Li}, {Li}, {Liang}, {Liao}, {Liu}, {Liu}, {Liu}, {Liu}, {Liu}, {Lu}, {Lu}, {Luo}, {Luo}, {Ma}, {Meng}, {Nang}, {Nie}, {Ou}, {Sai}, {Shang}, {Song}, {Sun}, {Tan}, {Tuo}, {Wang}, {Wang}, {Wang}, {Wang}, {Wang}, {Wang}, {Wen}, {Wu}, {Wu}, {Wu}, {Xiao}, {Xiao}, {Xiong}, {Xu}, {Yang}, {Yang}, {Yang}, {Yang}, {Yi}, {Yin}, {You}, {Zhang}, {Zhang}, {Zhang}, {Zhang}, {Zhang}, {Zhang}, {Zhang}, {Zhang}, {Zhang}, {Zhang}, {Zhang}, {Zhang}, {Zhang}, {Zhang}, {Zhang}, {Zhang}, {Zhao}, {Zhao}, {Zheng}, {Zheng}, {Zhou}, {Zhou}, {Zhu}, \& {Zhuang}}]{2020ApJ...902...18K}
{Kong}, L.~D., {Zhang}, S., {Chen}, Y.~P., {et~al.} 2020, \apj, 902, 18

\bibitem[{{Liao} {et~al.}(2020{\natexlab{a}}){Liao}, {Zhang}, {Chen}, {Zhang}, {Jin}, {Chang}, {Chen}, {Ge}, {Guo}, {Li}, {Li}, {Lu}, {Lu}, {Nie}, {Song}, {Yang}, {You}, {Zhao}, \& {Zhang}}]{2020JHEAp..27...24L}
{Liao}, J.-Y., {Zhang}, S., {Chen}, Y., {et~al.} 2020{\natexlab{a}}, Journal of High Energy Astrophysics, 27, 24

\bibitem[{{Liao} {et~al.}(2020{\natexlab{b}}){Liao}, {Zhang}, {Lu}, {Zhang}, {Li}, {Chang}, {Chen}, {Ge}, {Guo}, {Huang}, {Jin}, {Li}, {Li}, {Li}, {Liu}, {Lu}, {Nie}, {Song}, {Wang}, {You}, {Zhang}, {Zhao}, \& {Zhang}}]{2020JHEAp..27...14L}
{Liao}, J.-Y., {Zhang}, S., {Lu}, X.-F., {et~al.} 2020{\natexlab{b}}, Journal of High Energy Astrophysics, 27, 14

\bibitem[{{Liu, Q.} {et~al.}(2024){Liu, Q.}, {Santangelo, A.}, {Kong, L. D.}, {Ducci, L.}, {Ji, L.}, {Wang, W.}, {Serim, M. M.}, {Güngör, C.}, {Tuo, Y. L.}, \& {Serim, D.}}]{Liu_2024}
{Liu, Q.}, {Santangelo, A.}, {Kong, L. D.}, {et~al.} 2024, A\&A, 691, A215

\bibitem[{{Mandal} \& {Pal}(2022)}]{2022Ap&SS.367..112M}
{Mandal}, M. \& {Pal}, S. 2022, \apss, 367, 112

\bibitem[{{Mushtukov} {et~al.}(2015){Mushtukov}, {Suleimanov}, {Tsygankov}, \& {Poutanen}}]{2015MNRAS.447.1847M}
{Mushtukov}, A.~A., {Suleimanov}, V.~F., {Tsygankov}, S.~S., \& {Poutanen}, J. 2015, \mnras, 447, 1847

\bibitem[{{Postnov} {et~al.}(2015){Postnov}, {Gornostaev}, {Klochkov}, {Laplace}, {Lukin}, \& {Shakura}}]{2015MNRAS.452.1601P}
{Postnov}, K.~A., {Gornostaev}, M.~I., {Klochkov}, D., {et~al.} 2015, \mnras, 452, 1601

\bibitem[{{Raichur} \& {Paul}(2010)}]{2010MNRAS.406.2663R}
{Raichur}, H. \& {Paul}, B. 2010, \mnras, 406, 2663

\bibitem[{{Reig}(2011)}]{2011Ap&SS.332....1R}
{Reig}, P. 2011, \apss, 332, 1

\bibitem[{{Reig} \& {Nespoli}(2013)}]{2013A&A...551A...1R}
{Reig}, P. \& {Nespoli}, E. 2013, \aap, 551, A1

\bibitem[{{Serim} {et~al.}(2022){Serim}, {{\"O}z{\"u}do{\u{g}}ru}, {D{\"o}nmez}, {{\c{S}}ahiner}, {Serim}, {Baykal}, \& {{\.I}nam}}]{2022MNRAS.510.1438S}
{Serim}, M.~M., {{\"O}z{\"u}do{\u{g}}ru}, {\"O}.~C., {D{\"o}nmez}, {\c{C}}.~K., {et~al.} 2022, \mnras, 510, 1438

\bibitem[{{Tsygankov} {et~al.}(2017){Tsygankov}, {Wijnands}, {Lutovinov}, {Degenaar}, \& {Poutanen}}]{2017MNRAS.470..126T}
{Tsygankov}, S.~S., {Wijnands}, R., {Lutovinov}, A.~A., {Degenaar}, N., \& {Poutanen}, J. 2017, \mnras, 470, 126

\bibitem[{{Verner} {et~al.}(1996){Verner}, {Ferland}, {Korista}, \& {Yakovlev}}]{1996ApJ...465..487V}
{Verner}, D.~A., {Ferland}, G.~J., {Korista}, K.~T., \& {Yakovlev}, D.~G. 1996, \apj, 465, 487

\bibitem[{{Wang}(1995)}]{1995ApJ...449L.153W}
{Wang}, Y.~M. 1995, \apjl, 449, L153

\bibitem[{{Wilms} {et~al.}(2000){Wilms}, {Allen}, \& {McCray}}]{2000ApJ...542..914W}
{Wilms}, J., {Allen}, A., \& {McCray}, R. 2000, \apj, 542, 914

\bibitem[{{Wilson-Hodge} {et~al.}(2018){Wilson-Hodge}, {Malacaria}, {Jenke}, {Jaisawal}, {Kerr}, {Wolff}, {Arzoumanian}, {Chakrabarty}, {Doty}, {Gendreau}, {Guillot}, {Ho}, {LaMarr}, {Markwardt}, {{\"O}zel}, {Prigozhin}, {Ray}, {Ramos-Lerate}, {Remillard}, {Strohmayer}, {Vezie}, {Wood}, \& {NICER Science Team}}]{2018ApJ...863....9W}
{Wilson-Hodge}, C.~A., {Malacaria}, C., {Jenke}, P.~A., {et~al.} 2018, \apj, 863, 9

\bibitem[{{Zhang} {et~al.}(2020){Zhang}, {Li}, {Lu}, {Song}, {Xu}, {Liu}, {Chen}, {Cao}, {Bu}, {Chang}, {Chen}, {Chen}, {Chen}, {Chen}, {Chen}, {Cui}, {Cui}, {Deng}, {Dong}, {Du}, {Fu}, {Gao}, {Gao}, {Gao}, {Ge}, {Gu}, {Guan}, {Gungor}, {Guo}, {Han}, {Hu}, {Huang}, {Huo}, {Jia}, {Jiang}, {Jiang}, {Jin}, {Jin}, {Li}, {Li}, {Li}, {Li}, {Li}, {Li}, {Li}, {Li}, {Li}, {Li}, {Li}, {Liang}, {Liao}, {Liu}, {Liu}, {Liu}, {Liu}, {Liu}, {Liu}, {Lu}, {Lu}, {Luo}, {Ma}, {Meng}, {Nang}, {Nie}, {Ou}, {Qu}, {Sai}, {Shang}, {Shen}, {Sun}, {Tan}, {Tao}, {Tuo}, {Wang}, {Wang}, {Wang}, {Wang}, {Wang}, {Wang}, {Wang}, {Wen}, {Wu}, {Wu}, {Wu}, {Xiao}, {Xiong}, {Yan}, {Yang}, {Yang}, {Yang}, {Yi}, {Yuan}, {Zhang}, {Zhang}, {Zhang}, {Zhang}, {Zhang}, {Zhang}, {Zhang}, {Zhang}, {Zhang}, {Zhang}, {Zhang}, {Zhang}, {Zhang}, {Zhang}, {Zhang}, {Zhang}, {Zhang}, {Zhang}, {Zhang}, {Zhang}, {Zhao}, {Zhao}, {Zheng}, {Zhou}, {Zhu}, {Zhu}, {Zhuang}, \& {Insight-HXMT Team}}]{2020SCPMA..6349502Z}
{Zhang}, S.-N., {Li}, T., {Lu}, F., {et~al.} 2020, Science China Physics, Mechanics, and Astronomy, 63, 249502

\end{thebibliography}

\appendix
\onecolumn
\section{Additional material}

\begin{table*}[!htp]
    \centering
    \footnotesize
    \caption{Observation IDs as well as the best-fitting spectral parameters from Insight-HXMT during the 2021 outburst.}
    \label{tab:ObsIDs2021}
    \renewcommand\arraystretch{1.3}
    \setlength{\tabcolsep}{1.3mm}{
    \begin{tabular}{l|cccccccc}
    \hline \hline 
    \multirow{2}{*}{ObsID} & \multirow{2}{*}{Time Start [UTC]}  & \multirow{2}{*}{\thead{Exposure time [s] \\ (HE/ME/LE)}} & \multirow{2}{*}{MJD} & \multirow{2}{*}{Flux (10$^{-9}$ erg s$^{-1}$ cm$^{-1}$)} & \multirow{2}{*}{$\Gamma$} & \multirow{2}{*}{$E_{\rm cut}$} & \multirow{2}{*}{$\chi^2$/dof}\\ \\
    \hline 
P0304059001 & 2021-01-29 &      1820/3420/1029  & 59243.17 &    --   &  --                    &  --                     &  --      \\ 
P0304059002 & 2021-01-30 &      468/2040/721    & 59244.38 &    $1.79_{-0.01}^{+0.02}$ &       $0.18_{-0.05}^{+0.07}$ & $12.50_{-0.50}^{+1.15}$ &      714/786 \\ 
P0304059003 & 2021-01-31 &      2726/3270/1348  & 59245.24 &    $2.20_{-0.01}^{+0.01}$ &       $0.26_{-0.00}^{+0.12}$ & $13.28_{-0.00}^{+2.03}$ &      881/1068        \\ 
P0304059004 & 2021-02-01 &      1069/2310/585   & 59246.30 &    $2.00_{-0.01}^{+0.04}$ &       $0.40_{-0.09}^{+0.03}$ & $15.56_{-1.06}^{+1.04}$ &      611/725 \\ 
P0304059005 & 2021-02-02 &      2535/3269/1197  & 59247.16 &    $2.20_{-0.01}^{+0.01}$ &       $0.36_{-0.06}^{+0.05}$ & $15.14_{-0.79}^{+0.89}$ &      938/986 \\ 
P0304059006 & 2021-02-03 &      2830/3210/1380  & 59248.16 &    $2.15_{-0.01}^{+0.02}$ &       $0.32_{-0.04}^{+0.05}$ & $14.34_{-0.38}^{+0.91}$ &      825/1041        \\ 
P0304059007 & 2021-02-04 &      1282/2520/554   & 59249.22 &    $2.21_{-0.01}^{+0.02}$ &       $0.37_{-0.04}^{+0.05}$ & $15.48_{-0.42}^{+1.28}$ &      611/713 \\ 
P0304059008 & 2021-02-05 &      1195/2459/467   & 59250.21 &    $2.01_{-0.01}^{+0.03}$ &       $0.42_{-0.07}^{+0.05}$ & $16.54_{-0.74}^{+1.29}$ &      564/656 \\ 
P0304059009 & 2021-02-06 &      1177/2310/920   & 59251.20 &    $2.25_{-0.01}^{+0.03}$ &       $0.56_{-0.05}^{+0.03}$ & $21.15_{-1.07}^{+1.37}$ &      722/838 \\ 
P0304059010 & 2021-02-07 &      1396/2040/538   & 59252.31 &    $1.95_{-0.01}^{+0.02}$ &       $0.32_{-0.06}^{+0.05}$ & $14.32_{-0.63}^{+0.95}$ &      566/695 \\ 
P0304059011 & 2021-02-08 &      1002/2100/897   & 59253.17 &    $2.10_{-0.01}^{+0.03}$ &       $0.29_{-0.03}^{+0.09}$ & $14.44_{-0.08}^{+1.83}$ &      684/837 \\ 
P0304059012 & 2021-02-09 &      3138/2609/1915  & 59254.36 &    $2.21_{-0.01}^{+0.02}$ &       $0.40_{-0.06}^{+0.02}$ & $16.70_{-0.88}^{+0.55}$ &      976/1165        \\ 
P0304059013 & 2021-02-10 &      989/2550/1382   & 59255.24 &    $2.24_{-0.01}^{+0.02}$ &       $0.30_{-0.04}^{+0.03}$ & $15.31_{-0.51}^{+0.89}$ &      878/1042        \\ 
P0304059014 & 2021-02-10 &      2333/3120/1431  & 59256.07 &    $2.21_{-0.01}^{+0.02}$ &       $0.24_{-0.04}^{+0.05}$ & $14.00_{-0.37}^{+0.88}$ &      869/1054        \\ 
P0304059015 & 2021-02-12 &      749/3630/1589   & 59257.16 &    $2.11_{-0.01}^{+0.01}$ &       $0.25_{-0.02}^{+0.06}$ & $13.82_{-0.20}^{+1.28}$ &      844/1101        \\ 
P0304059016 & 2021-02-13 &      4365/4260/2034  & 59258.35 &    $2.23_{-0.01}^{+0.02}$ &       $0.29_{-0.02}^{+0.04}$ & $14.79_{-0.19}^{+0.74}$ &      1007/1197       \\ 
P0304059017 & 2021-02-14 &      2956/4260/1537  & 59259.21 &    $2.39_{-0.03}^{+0.04}$ &       $0.34_{-0.03}^{+0.03}$ & $15.82_{-0.31}^{+0.75}$ &      932/1099        \\ 
P0304059018 & 2021-02-16 &      5932/6869/1587  & 59261.93 &    $2.13_{-0.03}^{+0.04}$ &       $0.34_{-0.03}^{+0.02}$ & $15.51_{-0.41}^{+0.43}$ &      1014/1094       \\ 
P0304059019 & 2021-02-17 &      2159/6390/1570  & 59263.06 &    $2.29_{-0.02}^{+0.03}$ &       $0.37_{-0.03}^{+0.03}$ & $16.78_{-0.42}^{+0.70}$ &      1013/1099       \\ 
P0304059020 & 2021-02-18 &      2974/6750/957   & 59263.98 &    $2.10_{-0.02}^{+0.03}$ &       $0.29_{-0.04}^{+0.02}$ & $14.66_{-0.55}^{+0.41}$ &      891/871 \\ 
P0304059021 & 2021-02-19 &      1348/6180/724   & 59265.05 & --  &      --                     & --                      & --      \\ 
P0304059022 & 2021-02-20 &      4785/6929/1256  & 59265.90 &    $2.08_{-0.01}^{+0.02}$ &       $0.34_{-0.03}^{+0.03}$ & $14.70_{-0.40}^{+0.43}$ &      910/1004        \\ 
P0304059023 & 2021-02-21 &      2134/5520/882   & 59267.04 &    $2.01_{-0.02}^{+0.04}$ &       $0.37_{-0.04}^{+0.03}$ & $15.65_{-0.63}^{+0.67}$ &      727/821 \\ 
P0304059024 & 2021-02-22 &      2375/4920/415   & 59268.04 &    $1.85_{-0.02}^{+0.04}$ &       $0.32_{-0.05}^{+0.04}$ & $14.25_{-0.57}^{+0.64}$ &      529/654 \\ 
P0304059025 & 2021-02-23 &      2791/5580/1794  & 59269.03 &    $1.95_{-0.01}^{+0.02}$ &       $0.50_{-0.04}^{+0.02}$ & $18.16_{-0.85}^{+0.66}$ &      1068/1142       \\ 
P0304059026 & 2021-02-25 &      5832/6029/2762  & 59270.16 &    $1.97_{-0.01}^{+0.01}$ &       $0.46_{-0.03}^{+0.02}$ & $17.53_{-0.41}^{+0.63}$ &      1179/1278       \\ 
P0304059027 & 2021-02-25 &      3819/5699/837   & 59271.02 &    $1.65_{-0.01}^{+0.02}$ &       $0.40_{-0.06}^{+0.02}$ & $15.65_{-0.88}^{+0.40}$ &      731/763 \\ 
P0304059028 & 2021-02-26 &      1885/3450/1915  & 59271.60 &    $2.00_{-0.01}^{+0.01}$ &       $0.51_{-0.04}^{+0.03}$ & $17.85_{-0.73}^{+1.12}$ &      988/1155        \\ 
P0304059032 & 2021-02-28 &      4835/4649/2480  & 59273.20 &    $1.59_{-0.01}^{+0.01}$ &       $0.38_{-0.03}^{+0.03}$ & $15.40_{-0.50}^{+0.65}$ &      1084/1202       \\
P0304059033 & 2021-02-28 &      5396/5610/3011  & 59274.06 &    $1.50_{-0.01}^{+0.01}$ &       $0.40_{-0.04}^{+0.04}$ & $14.33_{-0.52}^{+0.72}$ &      971/1280        \\
P0304059034 & 2021-03-01 &      4693/5490/2656  & 59275.05 &    $1.67_{-0.01}^{+0.01}$ &       $0.52_{-0.04}^{+0.02}$ & $18.74_{-0.80}^{+0.66}$ &      1114/1227       \\
P0304059035 & 2021-03-02 &      5188/5400/2659  & 59276.05 &    $1.47_{-0.02}^{+0.03}$ &       $0.59_{-0.07}^{+0.01}$ & $20.25_{-1.45}^{+0.38}$ &      1047/1226       \\
P0304059036 & 2021-03-04 &      4086/3930/2513  & 59277.17 &    $1.37_{-0.01}^{+0.01}$ &       $0.47_{-0.07}^{+0.02}$ & $15.83_{-1.04}^{+0.60}$ &      984/1200        \\
P0304059037 & 2021-03-04 &      5234/5370/2811  & 59278.03 &    --  &   --                     & --                      & --      \\
P0304059038 & 2021-03-05 &      5461/6990/3897  & 59278.93 &    $1.16_{-0.01}^{+0.02}$ &       $0.48_{-0.04}^{+0.03}$ & $15.99_{-0.72}^{+0.88}$ &      1060/1320       \\
P0304059039 & 2021-03-06 &      5346/5460/2742  & 59279.95 &    $1.07_{-0.01}^{+0.01}$ &       $0.38_{-0.02}^{+0.08}$ & $13.55_{-0.09}^{+1.54}$ &      995/1233        \\
P0304059040 & 2021-03-07 &      2377/4890/3451  & 59280.81 &    $1.10_{-0.01}^{+0.01}$ &       $0.39_{-0.08}^{+0.04}$ & $13.89_{-0.63}^{+1.08}$ &      899/1303        \\
P0304059041 & 2021-03-08 &      4325/4260/3819  & 59281.53 &    --  &   --                     & --                      & --      \\
P0304059042 & 2021-03-09 &      4548/4140/3471  & 59282.52 &    $1.11_{-0.02}^{+0.01}$ &       $0.49_{-0.02}^{+0.06}$ & $14.04_{-0.33}^{+1.21}$ &      1117/1327       \\
P0304059043 & 2021-03-10 &      4609/4020/2682  & 59283.52 &    $1.02_{-0.01}^{+0.01}$ &       $0.58_{-0.05}^{+0.06}$ & $20.54_{-0.86}^{+2.42}$ &      921/1156        \\
P0304059044 & 2021-03-11 &      2724/2760/1227  & 59284.87 &    $0.84_{-0.01}^{+0.02}$ &       $0.59_{-0.08}^{+0.07}$ & $17.80_{-1.46}^{+2.30}$ &      731/856 \\
P0304059045 & 2021-03-12 &      4814/4920/2796  & 59285.91 &    $0.75_{-0.01}^{+0.01}$ &       $0.44_{-0.06}^{+0.04}$ & $13.65_{-0.77}^{+1.17}$ &      916/1205        \\
P0304059046 & 2021-03-14 &      2459/3300/1855  & 59287.36 &    $0.80_{-0.01}^{+0.01}$ &       $0.46_{-0.09}^{+0.10}$ & $17.91_{-1.68}^{+2.73}$ &      929/1015        \\ 
    \hline
    \end{tabular}
    }
\tablefoot{
The observations P0304059001, P0304059021, P0304059037, and P0304059041 do not yield an acceptable fit. \\
}

\end{table*}

\begin{table*}[htp]
\centering
\caption{Best-fitting spectral parameters of 2S 1417-624 in the hard X-ray bands from 2–140 keV during the 2018 outburst.}
\label{tab:ObsIDs}

\renewcommand\arraystretch{1.5}
\setlength{\tabcolsep}{2.5mm}{
\begin{tabular}{ll|cccccc}
\hline
 Number & ObsID & $\Gamma$ & $E_{\rm cut}$  & $\sigma_{\rm Fe}$ & kT$_{\rm bb}$ & Flux & $\chi^2$/dof \\ 
\hline 
1  &    P011470902701 & $0.42_{-0.04}^{+0.05}$ & $15.92_{-0.52}^{+1.32}$ & -- & -- & $2.06_{-0.01}^{+0.02}$ & 761/804 \\ \hline
2  &    P011470902901 & $0.46_{-0.06}^{+0.05}$ & $16.18_{-0.77}^{+1.24}$ & -- & -- & $2.08_{-0.01}^{+0.02}$ & 657/739 \\ \hline
3  &    P011470903001 & $0.35_{-0.04}^{+0.05}$ & $14.00_{-0.76}^{+0.52}$ & -- & -- & $2.11_{-0.03}^{+0.01}$ & 709/790 \\ \hline
4  &    P011470902601 & $0.35_{-0.05}^{+0.02}$ & $15.15_{-0.97}^{+0.13}$ & -- & -- & $2.17_{-0.01}^{+0.01}$ & 689/831 \\ \hline
5  &    P011470902801 & $0.61_{-0.07}^{+0.02}$ & $21.46_{-1.52}^{+1.02}$ & -- & -- & $2.29_{-0.01}^{+0.02}$ & 757/780 \\ \hline
6  &    P011470902201 & $0.48_{-0.04}^{+0.04}$ & $17.73_{-0.57}^{+0.92}$ & -- & -- & $2.47_{-0.04}^{+0.06}$ & 939/874 \\ \hline
7  &    P011470902501 & $0.46_{-0.04}^{+0.02}$ & $18.38_{-0.85}^{+0.43}$ & -- & -- & $2.51_{-0.01}^{+0.02}$ & 931/1059 \\ \hline
8  &    P011470902301 & $0.48_{-0.05}^{+0.03}$ & $19.05_{-0.91}^{+1.20}$ & -- & -- & $2.51_{-0.01}^{+0.04}$ & 563/691 \\ \hline
9  &    P011470900101 & $0.34_{-0.03}^{+0.01}$ & $18.09_{-0.47}^{+0.42}$ & -- & -- & $4.13_{-0.01}^{+0.03}$ & 981/1215 \\ \hline
10 &    P011470900102 & $0.32_{-0.04}^{+0.01}$ & $18.03_{-0.65}^{+0.27}$ & -- & -- & $4.19_{-0.01}^{+0.03}$ & 890/1070 \\ \hline
11 &    P011470900201 & $0.21_{-0.04}^{+0.04}$ & $15.80_{-0.52}^{+0.58}$ & -- & -- & $4.29_{-0.02}^{+0.05}$ & 706/770 \\ \hline
12 &    P011470900301 & $0.20_{-0.03}^{+0.02}$ & $16.31_{-0.34}^{+0.35}$ & $0.20_{-0.03}^{+0.12}$ & $0.55_{-0.09}^{+0.05}$ & $5.05_{-0.01}^{+0.01}$ & 1106/1346 \\ \hline
13 &    P011470900401 & $0.24_{-0.03}^{+0.01}$ & $17.34_{-0.39}^{+0.24}$ & $0.20_{-0.07}^{+0.31}$ & $0.51_{-0.06}^{+0.06}$ & $5.06_{-0.02}^{+0.04}$ & 1059/1256 \\ \hline
14 &    P011470900501 & $0.20_{-0.02}^{+0.01}$ & $16.52_{-0.24}^{+0.28}$ & -- & $0.30_{-0.03}^{+0.03}$ & $5.39_{-0.02}^{+0.04}$ & 1088/1164 \\ \hline
15 &    P011470900601 & $0.24_{-0.03}^{+0.02}$ & $17.28_{-0.40}^{+0.31}$ & -- & $0.45_{-0.21}^{+0.02}$ & $5.84_{-0.01}^{+0.03}$ & 1136/1182 \\ \hline
16 &    P011470900701 & $0.15_{-0.03}^{+0.02}$ & $15.89_{-0.30}^{+0.34}$ & $0.17_{-0.07}^{+0.11}$ & $0.57_{-0.03}^{+0.04}$ & $6.29_{-0.01}^{+0.02}$ & 1102/1349 \\ \hline
17 &    P011470900801 & $0.20_{-0.04}^{+0.01}$ & $16.76_{-0.50}^{+0.29}$ & $0.06_{-0.05}^{+0.03}$ & $0.51_{-0.05}^{+0.04}$ & $6.37_{-0.01}^{+0.03}$ & 1058/1182 \\ \hline
18 &    P011470901001 & $0.14_{-0.03}^{+0.02}$ & $15.99_{-0.35}^{+0.18}$ & $0.11_{-0.06}^{+0.13}$ & $0.53_{-0.05}^{+0.03}$ & $6.57_{-0.01}^{+0.04}$ & 999/1206 \\ \hline
19 &    P011470900901 & $0.17_{-0.03}^{+0.03}$ & $16.38_{-0.31}^{+0.45}$ & -- & $0.59_{-0.09}^{+0.06}$ & $6.89_{-0.01}^{+0.04}$ & 984/1078 \\ \hline
20 &    P011470901201 & $0.21_{-0.03}^{+0.01}$ & $16.78_{-0.33}^{+0.20}$ & $0.08_{-0.04}^{+0.11}$ & $0.47_{-0.07}^{+0.04}$ & $7.29_{-0.02}^{+0.02}$ & 1067/1191 \\ \hline
21 &    P011470901401 & $0.09_{-0.03}^{+0.02}$ & $15.25_{-0.28}^{+0.29}$ & -- & $0.58_{-0.07}^{+0.02}$ & $7.35_{-0.01}^{+0.03}$ & 1080/1230 \\ \hline
22 &    P011470901701 & $0.23_{-0.02}^{+0.01}$ & $16.73_{-0.19}^{+0.24}$ & $0.16_{-0.06}^{+0.14}$ & $0.60_{-0.06}^{+0.04}$ & $7.36_{-0.01}^{+0.03}$ & 1148/1326 \\ \hline
23 &    P011470901301 & $0.15_{-0.04}^{+0.01}$ & $16.61_{-0.55}^{+0.14}$ & $0.20_{-0.07}^{+0.12}$ & $0.72_{-0.07}^{+0.03}$ & $7.42_{-0.01}^{+0.02}$ & 1126/1335 \\ \hline
24 &    P011470901101 & $0.17_{-0.02}^{+0.05}$ & $16.70_{-0.24}^{+0.64}$ & -- & $0.57_{-0.12}^{+0.04}$ & $7.53_{-0.03}^{+0.02}$ & 833/1035 \\ \hline
25 &    P011470901601 & $0.20_{-0.03}^{+0.03}$ & $17.58_{-0.50}^{+0.59}$ & $0.17_{-0.04}^{+0.14}$ & $0.72_{-0.11}^{+0.02}$ & $7.85_{-0.01}^{+0.03}$ & 1203/1253 \\ \hline
26 &    P011470901801 & $0.17_{-0.03}^{+0.02}$ & $15.91_{-0.30}^{+0.36}$ & -- & $0.50_{-0.10}^{+0.04}$ & $7.92_{-0.01}^{+0.04}$ & 978/981 \\ \hline
27 &    P011470902001 & $0.17_{-0.02}^{+0.03}$ & $16.27_{-0.19}^{+0.31}$ & $0.14_{-0.05}^{+0.15}$ & $0.63_{-0.06}^{+0.05}$ & $8.02_{-0.02}^{+0.02}$ & 1140/1340 \\ \hline
28 &    P011470901901 & $0.20_{-0.02}^{+0.03}$ & $16.57_{-0.24}^{+0.37}$ & -- & $0.51_{-0.06}^{+0.02}$ & $8.08_{-0.02}^{+0.02}$ & 1064/1195 \\ \hline
29 &    P011470902101 & $0.14_{-0.03}^{+0.03}$ & $16.14_{-0.32}^{+0.32}$ & $0.13_{-0.02}^{+0.36}$ & $0.62_{-0.08}^{+0.07}$ & $8.10_{-0.03}^{+0.03}$ & 1027/1127 \\ \hline
\end{tabular} 
}

\tablefoot{
The unabsorbed flux is given in units of $10^{-9}$ erg cm$^{-2}$ s$^{-1}$ (ObsID sorted in an ascending flux order). The parameters $E_{\rm cut}$ (cut-off energy), $\sigma_{\rm Fe}$ (the width of iron line), and kT$_{\rm bb}$ (the blackbody temperature) are in units of keV. Uncertainties are given at the 68\% confidence. \\
}
\end{table*}

\end{document}